\documentclass[twocolumn]{aastex631} 
\usepackage{amsmath}
\usepackage{CJKutf8}
\usepackage{xcolor}

\begin{document}

\begin{CJK*}{UTF8}{gbsn}

\title{Wind Roche-lobe Overflow in Low-Mass Binaries: Exploring the Origin of Rapidly Rotating Blue Lurkers}

\correspondingauthor{Meng Sun}
\email{meng.sun@northwestern.edu}

\author[0000-0001-9037-6180]{Meng Sun(孙萌)}
\affiliation{Center for Interdisciplinary Exploration and Research in Astrophysics (CIERA), Northwestern University, 1800 Sherman Ave,
Evanston, IL 60201, USA}

\author[0000-0003-1241-7615]{Sasha Levina}
\affiliation{Center for Interdisciplinary Exploration and Research in Astrophysics (CIERA), Northwestern University, 1800 Sherman Ave,
Evanston, IL 60201, USA}
\affiliation{Haverford College Department of Physics and Astronomy, 370 Lancaster Ave, Haverford, PA 19041, USA} 

\author[0000-0001-6692-6410]{Seth Gossage}
\affiliation{Center for Interdisciplinary Exploration and Research in Astrophysics (CIERA), Northwestern University, 1800 Sherman Ave,
Evanston, IL 60201, USA}

\author[0000-0001-9236-5469]{Vicky Kalogera}
\affiliation{Center for Interdisciplinary Exploration and Research in Astrophysics (CIERA), Northwestern University, 1800 Sherman Ave,
Evanston, IL 60201, USA}
\affiliation{Department of Physics \& Astronomy, Northwestern University, 2145 Sheridan Road, Evanston, IL 60208, USA}

\author[0000-0002-3944-8406]{Emily M. Leiner}
\affiliation{Center for Interdisciplinary Exploration and Research in Astrophysics (CIERA), Northwestern University, 1800 Sherman Ave,
Evanston, IL 60201, USA}
\affiliation{Department of Physics, Illinois Institute of Technology, Chicago, IL 60616, USA}

\author[0000-0002-3881-9332]{Aaron M.\ Geller}
\affiliation{Center for Interdisciplinary Exploration and Research in Astrophysics (CIERA), Northwestern University, 1800 Sherman Ave,
Evanston, IL 60201, USA}
\affiliation{Department of Physics \& Astronomy, Northwestern University, 2145 Sheridan Road, Evanston, IL 60208, USA}

\author[0000-0002-2077-4914]{Zoheyr Doctor}
\affiliation{Center for Interdisciplinary Exploration and Research in Astrophysics (CIERA), Northwestern University, 1800 Sherman Ave,
Evanston, IL 60201, USA}

\begin{abstract}

Wind Roche-Lobe Overflow (WRLOF) is a mass-transfer mechanism proposed by Mohamed and Podsiadlowski (2007) for stellar binaries wherein the wind acceleration zone of the donor star exceeds its Roche lobe radius, allowing stellar wind material to be transferred to the accretor at enhanced rates. WRLOF may explain characteristics observed in blue lurkers and blue stragglers. While WRLOF has been implemented in rapid population synthesis codes, it has yet to be explored thoroughly in detailed binary models such as \texttt{MESA} (a 1D stellar evolution code), and over a wide range of initial binary configurations. We incorporate WRLOF accretion in \texttt{MESA} 
to investigate wide low-mass binaries at solar metallicity. We perform a parameter study over the initial orbital period and stellar mass. 
In most of the models where we consider angular momentum transfer during accretion, the accretor is spun up to the critical (or break-up) rotation rate. Then we assume the star develops a boosted wind to efficiently reduce the angular momentum so that it could maintain a sub-critical rotation. Balanced by boosted wind loss, the accretor only gains $\sim 2\%$ of its total mass, but can maintain a near-critical rotation rate during WRLOF. Notably, the mass-transfer efficiency is significantly smaller than in previous studies in which the rotation of the accretor is ignored. We compare our results to observational data of blue lurkers in M67 and find that the WRLOF mechanism can qualitatively explain the origin of their rapid rotation, their location on the HR diagram and their orbital periods.
\end{abstract}


\section{Introduction} 
\label{sec:intro}

\begin{figure}[tp]
\centering
\includegraphics[scale=0.04,angle=0]{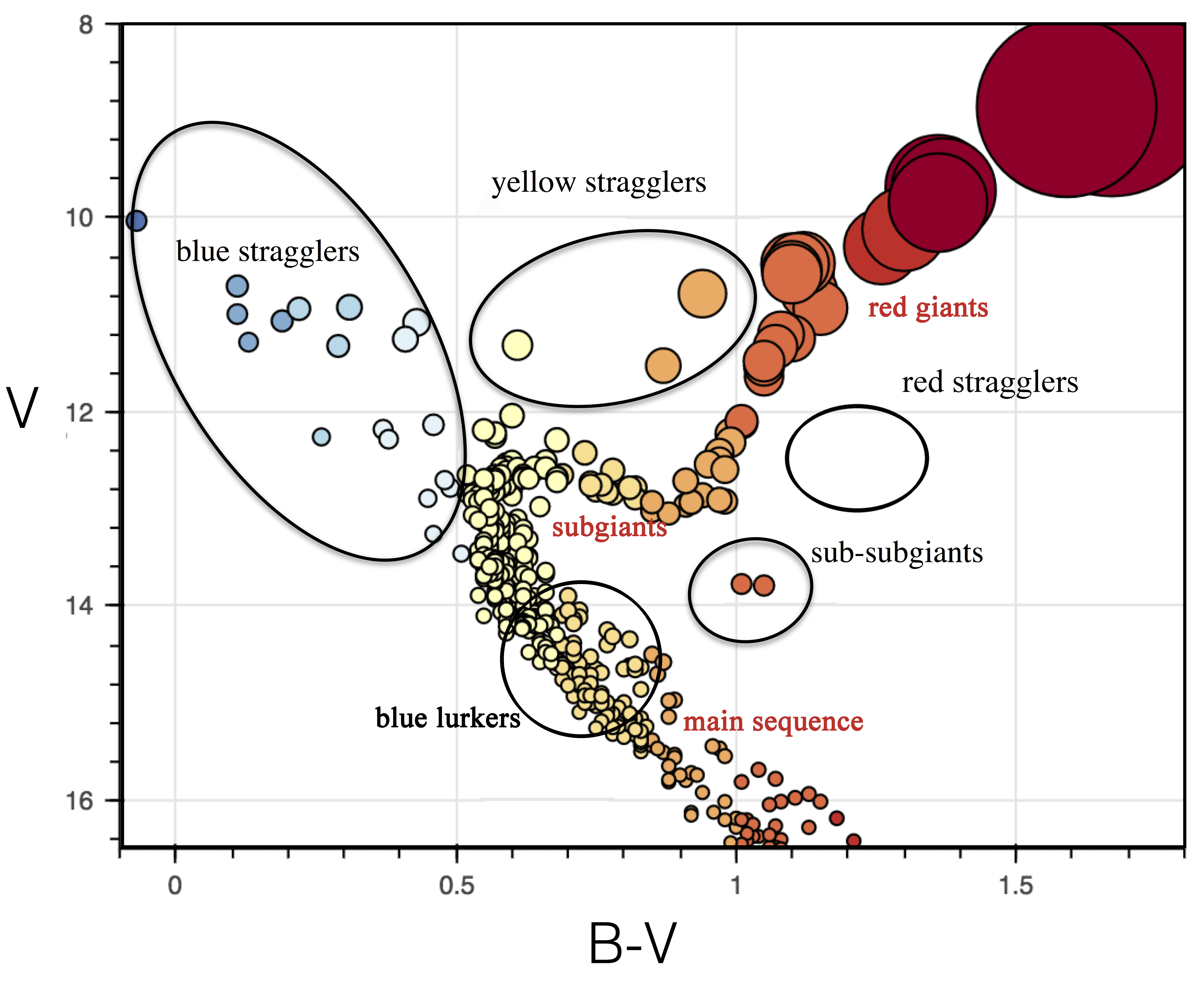}
\caption{A schematic color-magnitude diagram shows the relative locations of the various populations: blue stragglers, blue lurkers, yellow stragglers, red stragglers, and sub-subgiants (indicated with circles and black text), as well as key stellar evolutionary phases such as the main sequence, subgiant, and red giant branch (highlighted in red text). The base of the figure was provided by Emily Leiner through private communication at \url{https://sites.northwestern.edu/emilyleiner/research/}.
\label{CMD}}
\end{figure}

Observations of star clusters have revealed a population of stars that do not follow the expected single-star evolutionary pathways \citep{Sandage1953,Johnson1955,Ferraro1999}. These include yellow, red, and blue ``stragglers" and ``sub-subgiants". On the color-magnitude diagram of star clusters, yellow stragglers are located between the main-sequence (MS) turn-off and the red giant branch (RGB), but they are more luminous than stars in the subgiant phase \citep{Leiner2016}. Red stragglers are situated close to the RGB stars in color-magnitude space, but are redder than those on the RGB \citep{Albrow2001,Geller2017b,Stassun2023}, while blue stragglers stars (BSSs) are bluer or more luminous than the MS turn-off. Sub-subgiants are redder than the MS stars, but fainter than subgiants \citep{Belloni1998,Geller2017a,Leiner2017}. Figure \ref{CMD} provides the approximate locations of those ``outliers" on the stellar cluster's color-magnitude diagram. These stragglers provide excellent samples for exploring alternative stellar evolutionary pathways arising from: direct collision of two stars \citep{Hills1976,Hut1983,Leonard1989,Sills2001}, stellar merger \citep{Andronov2006,Ivanova2008,Knigge2009,Perets2009,Leigh2011,Naoz2014,Fragione2019}, and mass transfer (MT) in binary systems \citep{McCrea1964,Paczynski1971,Chen2004,Tian2006,Chen2008,Geller2011,Sun2021}.

Indeed, there is now mounting evidence that many of these outliers occur in binary systems. In the old (6-7 Gyr) open cluster NGC 188, which is a key cluster in the WIYN Open Cluster Survey (WOCS, \citealt{Mathieu2000}), \citet{Mathieu2009} discovered that 76\% of the BSSs are in binary systems with an orbital period smaller than 3000 days. In follow-up observations by \citet{Gosnell2014, Gosnell2015}, white dwarf (WD) companions were confirmed in seven binary systems containing a BSS, providing evidence many BSS are post-MT systems. These studies demonstrate that MT is the primary channel for forming BSSs in open clusters. On the BSS population side, it was investigated through the Cambridge STARS code \citep{2010Ap&SS.329..277C}, N-body simulation \citep{2013AJ....145....8G}, and the Binary Population and Spectral Synthesis code \citep{geinke2022}.


A new class of stars known as ``blue lurkers'' (BLs) has recently been identified by \citet{Leiner2019}. In the 4 Gyr-old open cluster M67, 11 MS stars are rotating anomalously rapidly with rotation periods of 2–8 days, while the solar-type MS stars in M67 have an average rotation period of 20–30 days. Moreover, these 11 BLs have a high binary fraction of (at least) 73\%, and 45\% of them have an orbital period less than 3000 days. \citet{Nine2023} confirmed the presence of WD companions in at least two of these blue lurker systems based on observed excess in HST/ACS far-UV photometry, indicating blue lurkers form via MT. Therefore, BLs are believed to be the low-mass, low-luminosity counterparts of BSSs.


In old open clusters (typically $>$ 1 Gyr), the majority of binaries containing a BSS or a BL have a wide orbit (with periods on the order of 1000 days, \citealt{Leiner2019,Leiner2021}). Regular Roche-Lobe Overflow (RLOF) occurs when the radius of the star is greater than its Roche-Lobe radius (which is defined in Equation (2) of \citet{1983ApJ...268..368E}). However, for instance, in the case of an orbit of 2000 days, it is unlikely for a star to fill its Roche-Lobe radius, even in the asymptotic giant branch (AGB) phase, unless the star is near a local maximum radius ($\sim 250 R_{\odot}$) during the thermal pulse phase \citep{Sun2023}. However, the thermal pulse phase is extremely short, and the maximum radius and number of pulses depend highly on the wind prescriptions. Therefore, a different mechanism for MT is needed to explain the presence of BSS and BL in wide binaries. One possible solution could be the Bondi-Hoyle-Lyttleton (BHL) wind transfer \citep{1939PCPS...35..405H,Bondi1944}. This process can only take place in the case of slow and dense, dusty winds, which are characteristic of AGB stars. Another proposed explanation is the so-called wind Roche-Lobe overflow (WRLOF) mechanism.

WRLOF occurs when the dust-driven winds of the primary star fill its Roche lobe and become focused in the direction of the secondary star \citep{Mohamed2007}. This WRLOF process results in greater accretion efficiency onto the secondary than the standard BHL wind transfer, where much material is lost from the system. Although originally proposed for slow dust-driven winds of AGB stars, WRLOF occurs more generally in systems where the outflows are accelerated over significant distances from the stellar surface, for example, in hot stars in high-mass X-ray binaries \citep{2019A&A...622L...3E,2021A&A...649L...2Z,2021ApJ...918...60W}. In another study, \citet{Saladino2019} proposed a different wind accretion prescription, for the accretion efficiency as a function of the ratio between wind and orbital speed, which is not limited to dust-driven wind.

WRLOF is relevant to various binary systems where standard mechanisms of MT (RLOF and BHL wind accretion) are insufficient to explain their observed characteristics. An example of such systems are carbon-enhanced metal-poor (CEMP) stars, which have a high frequency of about 20\% among the very metal-poor (VMP) stars with [Fe/H] $\leq -2.0$ \citep{Marsteller2005, Frebel2006, Lucatello2006}. \citet{Lucatello2005} found that all s-process-enhanced CEMP stars are in binary systems, indicating that these stars form via MT from an AGB primary. However, binary population synthesis models that use standard MT mechanisms predict a much lower CEMP/VMP ratio compared to observations unless (a) non-standard nucleosynthesis models are used or (b) there is a substantial change in the initial mass function. \citet{Abate2013} used a rapid binary population synthesis code (a combination of fitting functions from single-star evolutionary tracks and binary interaction prescriptions) implemented with WRLOF and narrowed the discrepancy between the observed CEMP population and theoretical estimations. WRLOF could also lead to possible Type Ia supernova events in wide Mira-type binaries \citep{2019MNRAS.485.5468I}.

\citet{Sun2023} used both WRLOF and regular RLOF to self-consistently model the evolution of WOCS 4540, a binary system consisting of a BSS and a WD in NGC 188. RGB and AGB wind was found to account for the majority of the MT, while regular RLOF only occurred during the largest thermal pulse of the donor in the thermal pulse AGB phase. Our analysis in this paper targets a much larger range of donor star masses, mass ratios and orbital separations, but follows a similar analysis for each individual model to investigate the impact of mass and angular momentum transfer via RLOF and WRLOF on the evolution of low-mass binaries (with donor mass range 0.9 - 8 $M_{\odot}$) and the formation and evolution of BLs.

In this paper, we implement the WRLOF prescription from \citet{Abate2013} into Modules for Experiments in Stellar Astrophysics (\texttt{MESA}, version 11701; \citealt{Paxton2011,Paxton2013,Paxton2015,Paxton2018,Paxton2019,Jermyn2022}) and evolve a grid of binaries. This is the first study implementing WRLOF in a large, self-consistent binary evolution grid to study its effects on binary evolution. In Section \ref{sec:WRLOF and others}, we provide background information on the WRLOF mechanism. Section \ref{sec:grid setting} describes the model settings in \texttt{MESA}, and we present the resulting grids and analyze an example single model in Section \ref{sec:results}. In Section \ref{sec:discussion}, we compare our results with observational BL data and discuss the effects of turning off angular momentum transfer and possible spinning down mechanisms. Finally, in Section \ref{sec:conclusions} we outline our main conclusions.

\section{Wind Roche-Lobe Overflow}
\label{sec:WRLOF and others}

WRLOF was first introduced in \citet{Mohamed2007} and \citet{Mohamed2010,Mohamed2010phd} in the context of explaining the observed mass outflow in the Mira binary: a wide, detached binary which consists of a cool AGB star and a WD. Neither regular RLOF nor BHL accretion could adequately explain this phenomenon. Beyond the Mira binary, this mechanism is applicable to a variety of systems containing an AGB star \citep{Abate2013,2019MNRAS.485.5468I,2020ApJ...900L..43L,2021ApJ...922...33R}.

The key factor in differentiating between traditional BHL accretion and the WRLOF mechanism is the speed of the wind $v_{\text{w}}$. If the wind speed is much higher than the orbital velocity $v_{\text{orb}}$ of the secondary star, wind MT will adhere to the BHL description. In contrast, in a closer binary containing an AGB star, wind MT will occur via the WRLOF mechanism when the wind velocity is lower than or comparable to the orbital speed of the secondary star. As described in \citet{Abate2013}, the stellar wind inside the wind acceleration zone can be efficiently transferred to the secondary star if the radius of this zone exceeds the Roche-lobe radius.



For the WRLOF accretion efficiency $\beta_{\text{WRLOF}}$, we apply the parabolic fitting function from \citet{Abate2013}, which converts the WRLOF accretion efficiency from numerical hydrodynamical simulations to a 1-D format, into our models:

\begin{equation}
    \beta_{\text{WRLOF}} = c_1 \left(\frac{R_{\text{d}}}{R_{\text{RL,1}}}\right)^2 + c_2 \left(\frac{R_{\text{d}}}{R_{\text{RL,1}}}\right) + c_3.
\end{equation} 
Here $R_{\text{RL,1}}$ is the Roche-lobe radius of the donor, and the parabolic fit parameters are $c_1 = -0.284$, $c_2 = 0.918$, and $c_3 = -0.234$. $R_{\text{d}}$ is the radius of the wind acceleration zone, which is defined by \citet{Honfer2007} as
\begin{equation}
    R_{\text{d}} = \frac{1}{2} R_{\ast} \left(\frac{T_{\text{eff}}}{T_{\text{cond}}}\right)^{2.5},
\end{equation} 
where $R_{\ast}$ is the stellar radius, $T_{\text{eff}}$ is the effective temperature of the star, and $T_{\text{cond}}$ is the condensation temperature of the dust in the wind. Following \citet{Abate2013}, we used the condensation temperature of $T_{\text{cond}}=1500$ K for carbon-rich dust (AGB stars with C/O $>$ 1) and $T_{\text{cond}}=1000$ K for oxygen-rich dust (AGB stars with C/O $<$ 1).

The $\beta_{\text{WRLOF}}$ term is a function of the mass-ratio $q=M_2/M_1$ between the accretor ($M_2$) and the donor ($M_1$). Throughout this paper, properties of the donor are assigned the subscript ``1'' and properties of the accretor have the subscript ``2''. According to the numerical simulations in \citet{Mohamed2010phd}, there is a maximum efficiency when $\beta_{\text{WRLOF,max}} = 0.5$. Based on the above constraints, the final form of the $\beta_{\text{WRLOF}}$ function is 

\begin{equation}
    \beta_{\text{WRLOF}} = \min\left\{\frac{25}{9} q^2 \left[c_1 x^2 + c_2 x + c_3\right], \beta_{\text{WRLOF,max}}\right\}.
\label{wrlof}
\end{equation}


Throughout this paper, we use the terms ``regular RLOF'' to refer to the traditional RLOF mechanism. The ``WRLOF'' is modeled using the prescription by \citet{Abate2013}. Generally, the efficiency of WRLOF lies between that of BHL accretion and regular RLOF, with regular RLOF being the most efficient type of MT. In our study, which focuses on low-mass binaries, we apply both WRLOF and regular RLOF to explore the parameter space using detailed binary evolution. We also discuss the effect of the traditional BHL accretion in Section \ref{sec:discussion:bhl}. Table \ref{table: MT Prescriptions} summarizes the MT prescriptions.

\begin{table*}[]
\caption{Mass Transfer Prescriptions}
\begin{center}
\begin{tabular}{c c c}
\hline\hline 
Name	& When it occurs & Efficiency \\
\hline
Regular RLOF & $R_{\ast} \gtrsim R_{\text{RL,1}}$ & Conservative mass transfer \\
WRLOF  & $R_{\rm d} \gtrsim R_{\text{RL,1}}$ & Equation \ref{wrlof}\\
BHL accretion  & $v_{\text{w}} \gg v_{\text{orb}}$ & Equation \ref{bhl}, but not included in the fiducial grid\\
\hline
\label{table: MT Prescriptions}
\end{tabular}
\end{center}
\end{table*}

\section{\texttt{MESA} Binary Grid Settings}
\label{sec:grid setting}

In this study, we assume that the two stars in a binary system are born and evolved simultaneously. The \texttt{MESA binary} module (version 11701) is used with the \texttt{evolve\_both\_stars=.true.} option applied. \texttt{MESA} performs 1-dimensional models of stellar structure, including a nuclear reaction network, boundary conditions, energy transportation, mixing, and rotation. We adopt solar metallicity $Z_{\odot} = 0.0142$ \citep{Asplund2009} 
throughout the parameter space. \texttt{MESA} does not include the WRLOF mechanism in the \texttt{binary} module by default, so we implement Equation \ref{wrlof} in our simulations. Our inlists and code is shared at \url{https://zenodo.org/records/11043393}. In addition to the \texttt{MESA} code, we utilize the framework of the \texttt{POSYDON} code (POpulation SYnthesis with Detailed Binary-evolution simulatiONs, \citealt{Fragos2022}), to initiate numerous binary grids across a wide parameter range.

\begin{table*}[]
\caption{Initial Binary Parameters and Steps}
\begin{center}
\begin{tabular}{c c c c c c}
\hline\hline 
$M_1/M_{\odot}$	& $\Delta \log(M_1/M_{\odot})$ & $M_2/M_{\odot}$ & $\Delta M_2/M_{\odot}$ & $P_{\rm orb}/{\rm day}$ & $\Delta \log (P_{\rm orb}/{\rm day})$	\\
\hline
0.9 - 8 & 0.056 & 0.9 - 1.3 & 0.1 & $10^2-2\times 10^5$ & 0.17 \\
\hline
\label{table: initial model}
\end{tabular}
\end{center}
\end{table*}

We run a $ N_{M_1}\times N_{M_2}\times N_{P_\mathrm{orb}} = (15\,\mathrm{to}\,18)\times 5\times 30$ grid, where $N$ stands for the number of initial values of $M_1$, $M_2$, and orbital period $P_{\rm{orb}}$. To ensure that the donor star evolves first, initial $M_1$ must always be greater than initial $M_2$ (which is also why the number of grid points for $M_1$ differs from that of $M_2$). 
The ranges and steps of the grid are shown in Table \ref{table: initial model}. We selected these parameters based on two criteria: (a) the donor star must evolve to an AGB phase; and (b) the initial $P_{\rm orb}$ must be wide enough to allow for a wind MT phase before the occurrence of case A or case B MT. 

We adopt most of the physics of single star and binary evolution described in \citealt{Fragos2022}, including stellar winds, MT, and tides. For mass accretion onto a non-degenerate star, the accretor gains mass as well as the angular momentum as described in \citet{deMink2013}, which causes the accretor to spin up. In most of our models, where binary interactions are treated self-consistently, angular momentum may be transferred efficiently via MT generally, whether it is regular RLOF or WRLOF (see Figure \ref{fig:rot} and Section \ref{subsec:Effects of Angular Momentum Transfer} for more details). As a result, the accretor can be easily spun up near the critical rotation rate, $\omega_{\rm crit}$. When this happens, we boost the stellar wind so that the rotation rate of the star $\omega_{\rm s}$ is lowered below the critical rotation rate (see Equation 1 in \citealt{Fragos2022}). In this work, our goal is to reproduce the features of the recently identified BLs, which demand less efficient accretion. To achieve this, we applied the \texttt{POSYDON} boosted wind approach (additional discussions can be found in \ref{subsec:possible spin-down}). The effectiveness of the enhanced wind at critical rotation in massive and low-mass binary systems remains uncertain, which also requires future investigation.

The termination conditions of the grid of models are similar to as described in \citet{Fragos2022}: a star's age exceeds 13.8 Gyr, a star evolves into a WD or reaches end of core C-burning, or unstable MT occurs followed by a common envelope (CE) phase. Due to numerical issues, 14\% of the models are terminated before meeting the termination conditions listed above. The majority of the not-converged models are at the donor's thermal pulse AGB phase, which challenges the code's convergence.

Furthermore, dynamically unstable MT (typically associated with a large MT rate) results in a CE phase. The 1-D stellar structure \texttt{MESA} code is not suitable for modeling such an event in detail. In \citet{Fragos2022}, the first version of \texttt{POSYDON} focuses on massive binaries, the criterion for terminating the code is set to a MT rate of $0.1\,M_\odot$ $\text{yr}^{-1}$. However, since we are considering low-mass binary grids, the criterion for unstable MT should be lower. Initially, we applied the same termination criterion as in \citet{Fragos2022}, but we found a significant number of models that crashed next to the island of the stable MT cases. These crashed models share common characteristics, such as $q> 1.5$, a rapid increase in MT rate over a very short timescale (shorter than star's nuclear and thermal timescale), and a MT rate exceeding $10^{-6}\,M_\odot$ $\text{yr}^{-1}$ in the last step. These are all indicative of dynamically unstable MT. Therefore, the definition of unstable MT models in this work differs from that of \citet{Fragos2022}. Our criterion for unstable MT is a MT rate exceeding $10^{-6}\,M_\odot$ $\text{yr}^{-1}$, but only if a model is not converged at the last evolution step. Additionally, using $10^{-6}\,M_\odot$ $\text{yr}^{-1}$ helps correctly classify most of the dynamically unstable MT models (i.e., using a bigger rate such as $10^{-5}\,M_\odot$ $\text{yr}^{-1}$ still results in a small number of non-converged models, which should be the unstable MT models). This criterion for unstable MT applies to both regular RLOF and WRLOF.

In rapid binary population synthesis codes, dynamically unstable MT is usually considered when the mass ratio between the donor and accretor exceeds a critical value, denoted as $q_{\rm crit}$. This critical value is determined by a parameterized equation as a function of the stellar global physical quantities. \citet{2011ApJ...739L..48W} argued that the $q_{\rm crit}$ criterion is too strict to accurately determine an unstable MT process. They also suggested that the exact response of the donor star to mass loss should be self-consistently simulated using a detailed evolutionary code. Consequently, the radius and MT rate could be accurately calculated. This represents a key difference between our work and other rapid binary population synthesis codes implemented with WRLOF.

When the donor star enters the thermal pulse AGB phase, following the \texttt{MIST} project \citep{Dotter2016,Choi2016}, we switched the wind prescription to the Bl\"{o}cker scheme. The thermal pulse AGB phase is defined as the helium core mass being greater than the carbon core mass by less than 0.1 $M_{\odot}$ and the central helium of the star being depleted. In \texttt{MESA}'s definition, for instance, the helium core mass is defined where the $^{4}{\rm He}$ mass fraction in a zone is above 0.01, and the hydrogen mass fraction is smaller than 0.01. This stronger wind scheme significantly improves the convergence issue when modeling the thermal pulse AGB phase with mass transfer and provides a reasonable initial and final mass relation for single star evolution.

For simplification, we only allow wind MT from the donor star to the accretor. In reality, at a later stage, the accretor evolves, and its stellar wind becomes non-negligible. When the accretor enters an AGB phase, the dust-driven wind could transfer back to the donor star through this WRLOF prescription. The original accretor could end up being less massive than our simulation. Furthermore, we treat the orbital eccentricity as zero. In an eccentric orbit, the resulting phase-averaged Roche-lobe radius could lead to a different MT rate through regular RLOF and WRLOF. We recognize the importance of both and plan to implement reverse MT from the accretor and non-zero eccentricity in future work. 

\section{Results of Binary Evolution with WRLOF}
\label{sec:results}

In this section, we show the resulting grid of models, with and and without the application of the WRLOF mechanism. Following a discussion of the grids, we select an example model with WRLOF and conduct a detailed analysis of its evolution. 

\subsection{The Fiducial Grid of Binary Evolution Models}

\begin{figure*}[t!]
\centering
\includegraphics[scale=0.25,angle=0]{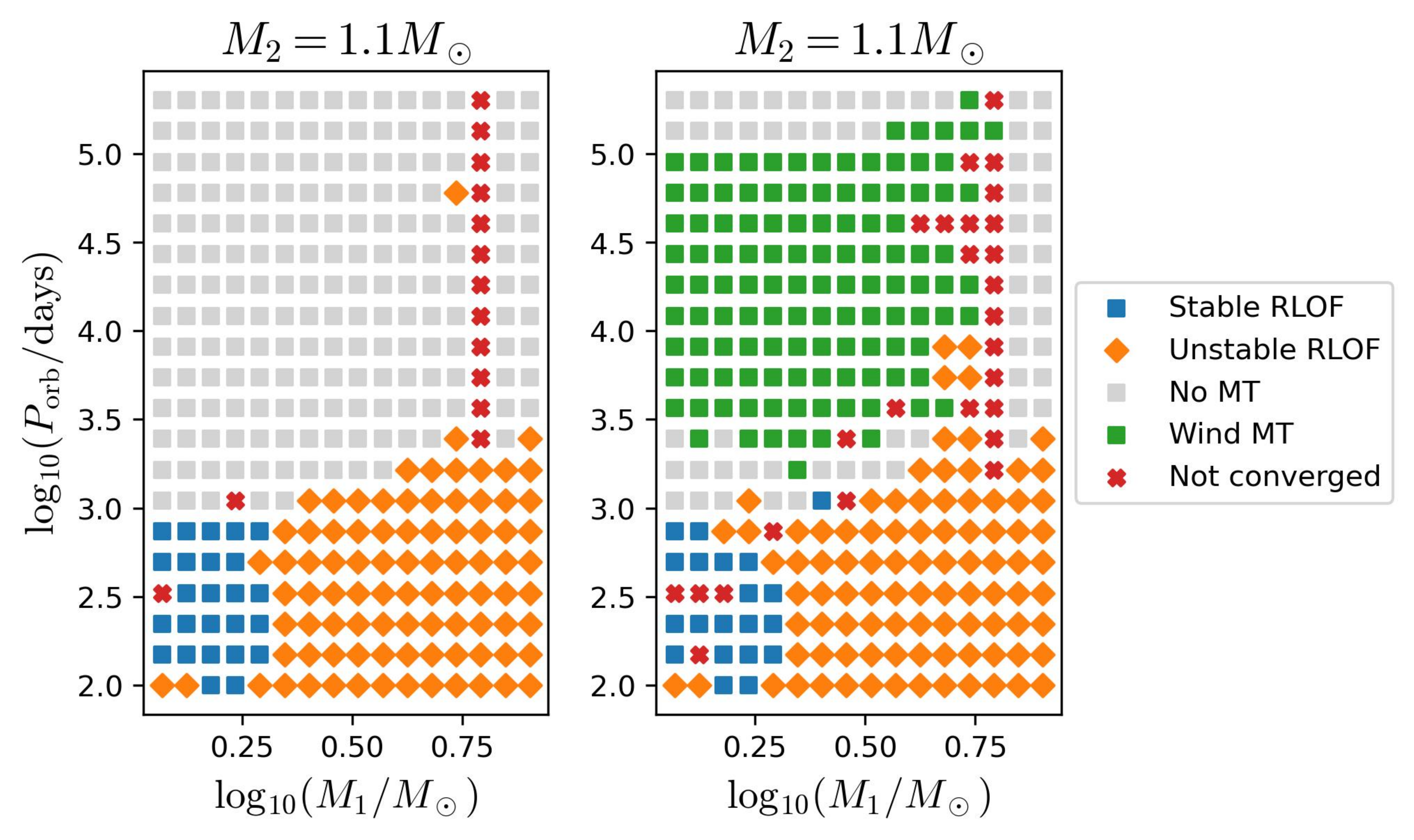}
\caption{A slice of the grid with initial $M_2 = 1.1 M_\odot$. The marker shape and color indicate the termination condition for each model as indicated in the legend. The axes are the initial parameters for $M_1$ and $P_{\text{orb}}$ on a logarithmic scale. The models in the left panel only include MT via RLOF whereas the right includes MT via both RLOF and WRLOF.}
\label{fig:end}
\end{figure*}

Using the pre-calculated zero-age main-sequence (ZAMS) models described in Section 5.1 of \citet{Fragos2022}, we show the grids of binary evolution models as two-dimensional slices in Figure \ref{fig:end}. In this figure, the initial secondary star mass is fixed at $M_2=1.1\,M_{\odot}$. The horizontal axis of each subplot is the initial donor mass, ranging from 1.17 $M_\odot$ to 8.0 $M_\odot$. The vertical axis is the initial orbital period ranging from $10^2$ to $2 \times 10^5$ days. The final status of each model is represented with either a blue square (stable regular RLOF), green square (WRLOF), orange diamond (unstable MT), or gray square (no MT). 

The left panel of Figure \ref{fig:end} shows the models with WRLOF turned off. The right panel shows the models with WRLOF applied. 
The models which did not converge are shown as red crosses. While a greater number of WRLOF models ended due to convergence issues (roughly 14\% vs. 4\%), these convergence issues do not affect the main results as the not-converged models were not used for further analysis.

In the left panel of Figure \ref{fig:end}, for donor mass smaller/bigger than $1.8\,M_{\odot}$ ($\log_{10}(M_1/M_\odot)=0.26$), no RLOF models have an initial period $\gtrsim 1000/3000$ days, respectively. Stable RLOF models occupy a small island at the bottom left corner on each panel, where the initial mass ratio of the binaries is close to one. For a mass ratio smaller than $q = 0.62$, the system undergoes unstable RLOF, leading to a CE phase where the evolution is terminated. As shown in the right panel of Figure \ref{fig:end}, systems that undergo WRLOF have initial $M_1$ ranging from 1.17 to 5.45 $M_\odot$ and initial $P_{\rm orb}$ ranging from $2.45 \times 10^3$ to $1.34 \times 10^5$ days.

\begin{figure*}[b!]
\centering
\includegraphics[scale=0.55,angle=0]{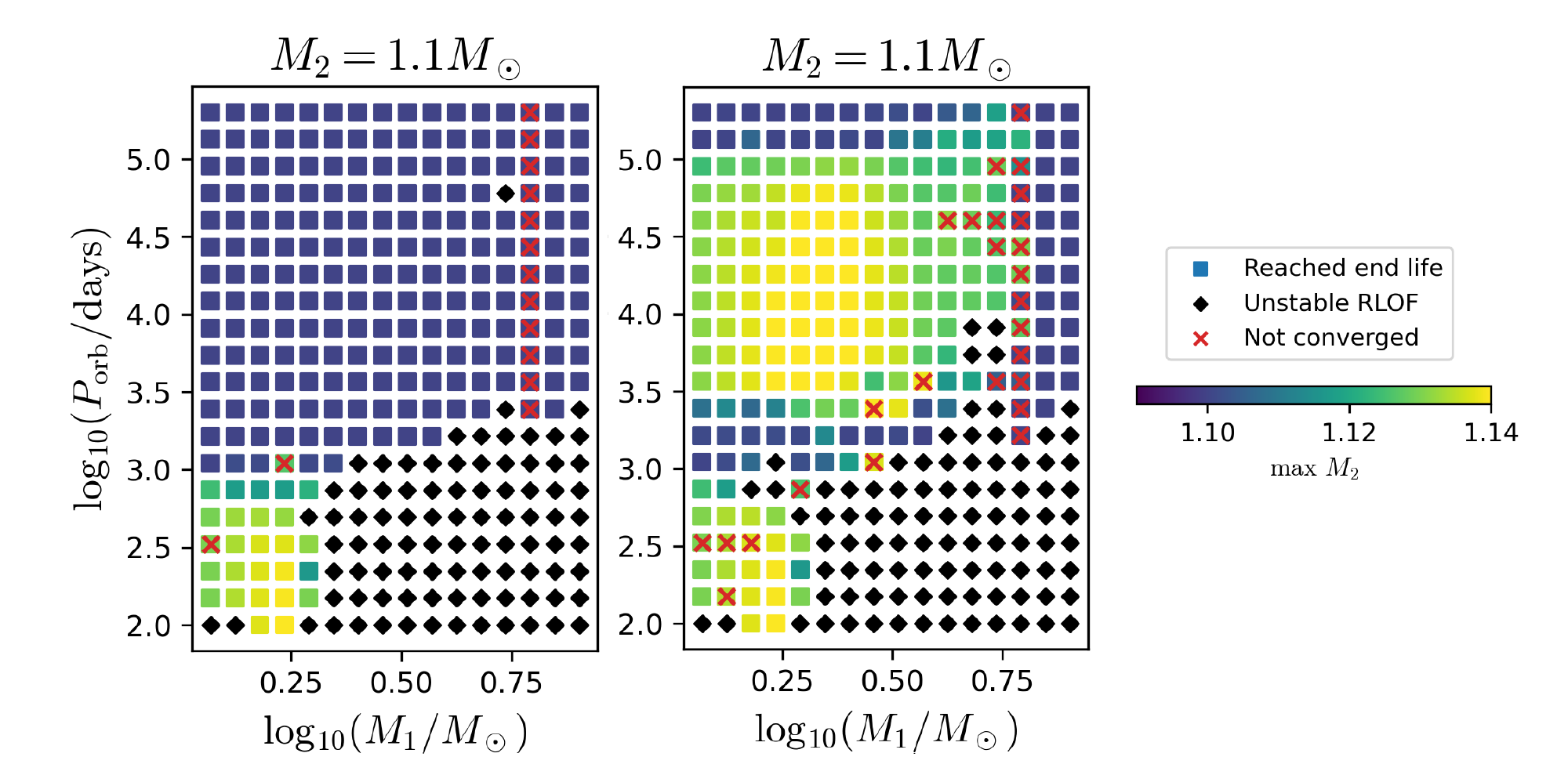}
\caption{A slice of the grid with initial $M_2 = 1.1 M_\odot$, where the plot on the left if for models that include MT via RLOF only, and the models on the right have both RLOF and WRLOF applied. The marker shape shows the termination condition for each model and the color indicates the accretor mass after MT, $M_{\rm 2,f}$. The axes are the initial parameters for $M_1$ and $P_{\text{orb}}$ on a logarithmic scale. \label{fig:m2}}
\end{figure*}

In figure \ref{fig:m2} we display a comparison between the final accretor mass for the same two grid slices as in Figure \ref{fig:end}. The termination conditions are indicated using the marker shape. Stable MT and no MT cases are shown as squares, unstable MT models are indicated with diamonds, and not-converged models are covered with red crosses. The color scale shows the final mass of the accretor ($M_{\rm 2,f}$). In the bottom left corner of both plots, the accretor gains $\sim 0.04\,M_\odot$ on average from regular case-B MT. Without wind MT, the accretors in the no-MT island maintain a nearly constant mass, as the accretor is almost un-evolved and so does not develop strong winds. Therefore, $M_2$ remains constant and evolves as a single MS star without interacting with the other star in the system.

With WRLOF turned on, as in the right panel of Figure \ref{fig:m2}, a significant number of models in a wide orbit gain mass. Near initial $M_1\sim 2.5\,M_{\odot}$ and initial $P_{\rm orb}\sim 10^4$ days, the accretor gains a maximum mass of 0.05 $M_{\odot}$ from the donor's wind. The accretion is reduced by the boosted wind when the star reaches the critical rotation rate, $\omega/\omega_{\rm crit}\simeq 1$. Not all accretors can gain 0.05 $M_{\odot}$ of material from the wind of the donor, though. Over the whole grid, the accretor mass increases by about 0.04 $M_{\odot}$ ($\sim$ 2\% of its total mass) on average through WRLOF. 

In summary, the place where WRLOF is having an effect is at higher orbital periods. The accretor can gain $\sim 0.04\,M_\odot$ from either regular RLOF (bottom left corner) or WRLOF (upper-left region of right panel). The average mass gain is similar for other initial accretor masses listed in Table \ref{table: initial model}.

\begin{figure*}[h]
\centering
\includegraphics[scale=0.55,angle=0]{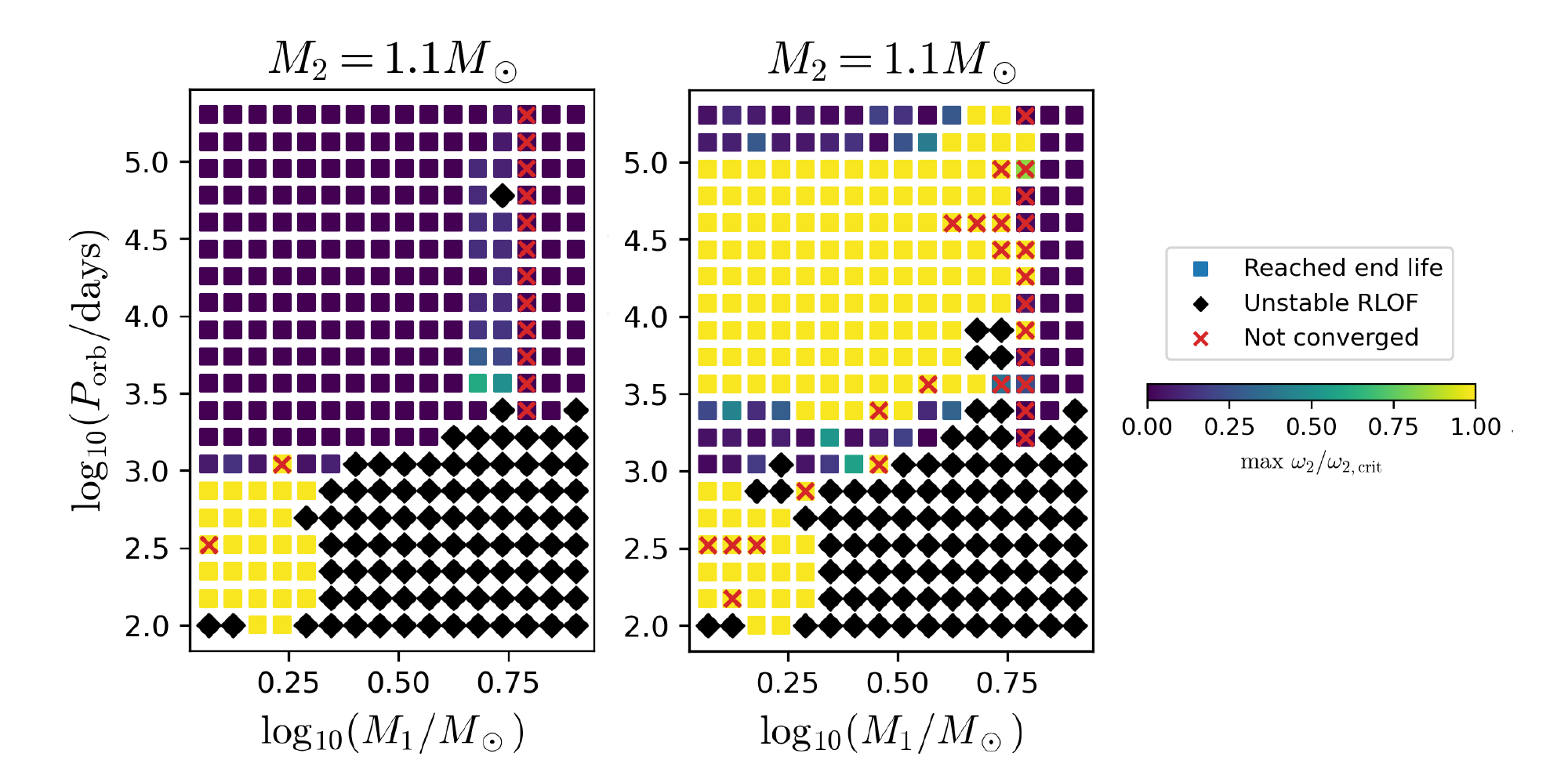}
\caption{The same as Figure \ref{fig:m2}, but where the color indicates the maximum surface rotation speed of the accretor as a fraction of its critical rotation speed, $\omega/\omega_\text{crit}$. }
\label{fig:rot}
\end{figure*}

Using the same initial parameter space as Figure \ref{fig:end}, Figure \ref{fig:rot} describes the distribution of the maximum rotation rate of the accretor. The color in Figure \ref{fig:rot} shows the maximum surface rotation rate of the accretor as a fraction of its critical rotation during the evolution. The plots are otherwise analogous to Figure \ref{fig:m2}. The subplot on the right illustrates that when the accretor gains mass, its surface rotational speed increases and can reach the critical rate, even though the regular RLOF and WRLOF efficiency of the accretion may be small (considering that only 0.04 $M_{\odot}$ is accreted). The angular momentum transfer through regular RLOF and WRLOF are both efficient (See Section \ref{subsec:Effects of Angular Momentum Transfer} for further discussion). Near the end of the simulation, accretors from post-regular RLOF and WRLOF cases spin down due to their evolution and expansion in radius, causing $\omega/\omega_\text{crit}$ to drop below 0.5 for some models.

All models begin with a wide orbit in order to allow the donor star to evolve into an AGB phase and develop slow winds. At this point, the system's orbital angular momentum is much greater than the stars' individual angular momenta. The angular momentum evolution of the two cases differs in that when WRLOF is applied, part of the angular momentum from the donor's wind transfers onto the accretor. In the case without WRLOF, that angular momentum instead escapes from the system. This angular momentum is much smaller than the orbital angular momentum of the system and thus the final orbital period distribution with and without WRLOF does not differ significantly. For other initial accretor masses, analogous plots to Figure \ref{fig:end} to Figure \ref{fig:rot} give very similar results. Therefore, we only use $M_2 = 1.1$ $M_{\odot}$ in the grid analysis here.

\subsection{Evolution of an Example WRLOF Binary}
\label{sec:fiducial detailed model}

As an example, we show here a detailed analysis of one of the binary systems simulated in our grids with WRLOF applied. We select a converged model which terminated due to the donor reaching the end of its life as a WD, and in which the mass of the accretor increased. The example model has initial parameters $M_1 = 1.33 M_\odot, M_2 = 1.10 M_\odot$, and $P_{\text{orb}} = 3661$ days. 

\begin{figure}[tp]
\centering
\includegraphics[scale=0.47,angle=0]{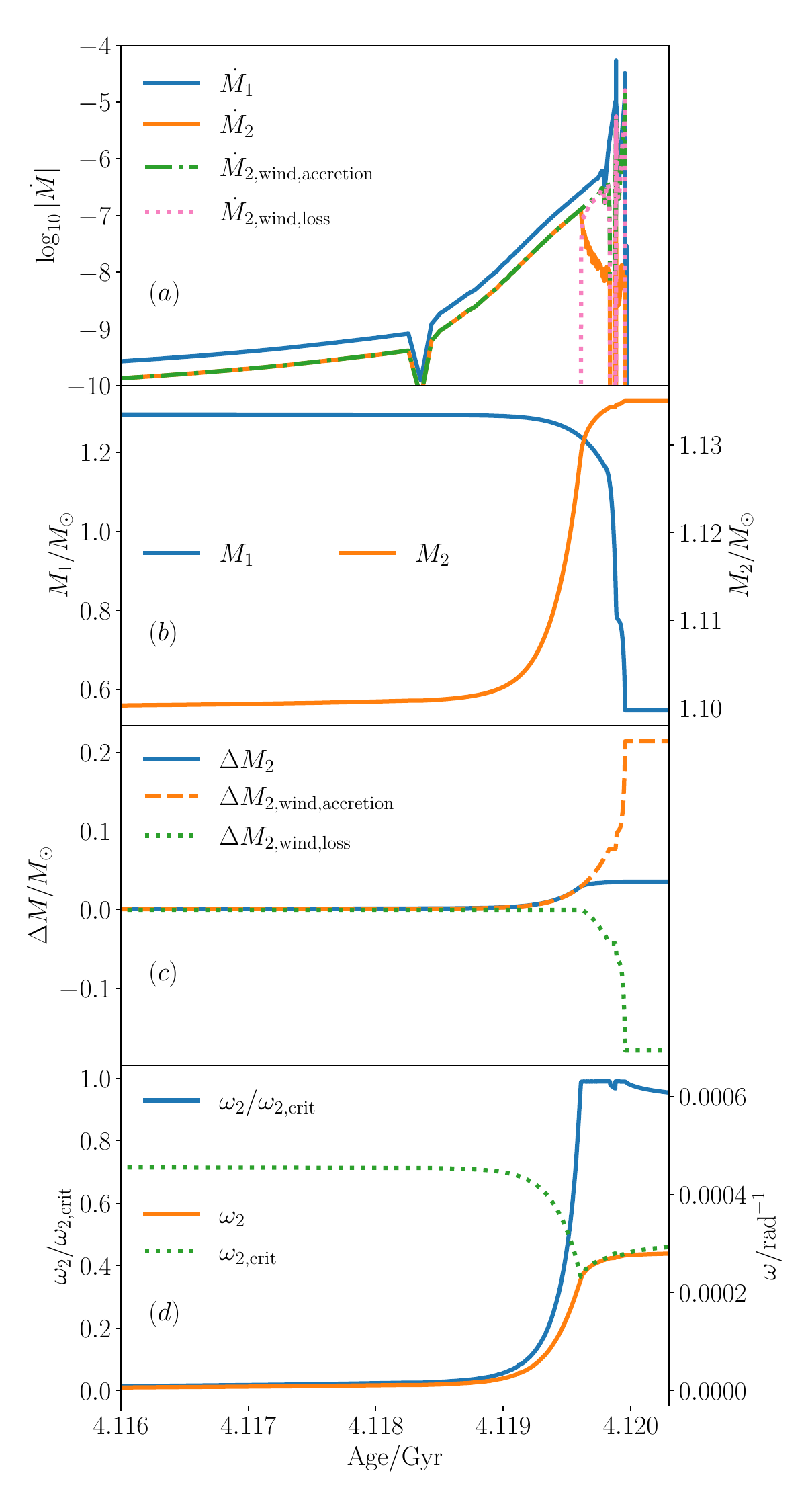} 
\caption{The WRLOF stage from one of the models in the fiducial grid during the AGB phase of the donor star, all as a function of the age of the system. Panel (a) shows the total mass change rate of the donor (blue) and the accretor (orange) and the mass-accretion rate (green) and mass loss due to wind (pink) of the accretor; Panel (b) shows the mass of the donor (blue) and the accretor (orange); Panel (c) shows the total change in mass of the accretor (blue), the change in mass of the accretor due to wind accretion (orange), and the change in mass of the accretor due to mass loss (green); Panel (d) shows the surface rotation speed of the accretor as a fraction of critical rotation (blue), the surface rotation speed of the accretor (orange), and the critical rotation speed of the accretor (green). This model has the initial settings of $M_1=1.33$ $M_{\odot}$, $M_2=1.1$ $M_{\odot}$, and $P_{\rm orb}=3661$ days.}
\label{fig:Mdot_t_wRLOF}
\end{figure}

In Figure \ref{fig:Mdot_t_wRLOF} we show the binary properties during the WRLOF stage of the donor star's AGB phase. In the top panel, the total mass change rate from the donor star, $\dot{M}_1$, is represented by a blue solid line, while that of the accretor, $\dot{M}_2$, is shown by an orange solid line. The mass loss of the donor star is completely through the stellar wind. As this binary has initial $P_{\rm orb} = 3661$ days, the system has no chance of going through regular RLOF (initial $R_{\rm RL,1}\sim 530R_{\odot}$ and the maximum radius of the donor is $\sim 200 R_{\odot}$). We also ignore the accretor's wind transfer back to the donor. Therefore, the total mass change of the donor star is equivalent to the wind mass-loss rate from the donor $\dot{M}_{1,\,{\rm wind}}$. The total mass change of the accretor has two sources: (a) accretion from the donor's wind, $\dot{M}_{2,\,{\rm wind,\,accretion}}$ (green dotted-dash line), and (b) mass-loss through its own wind, $\dot{M}_{2,\,{\rm wind,\,loss}}$ (pink dotted line). 

In summary, the relations between the mass change rates described in Figure \ref{fig:Mdot_t_wRLOF} top panel are:
\begin{align*}
&\dot{M}_1 = \dot{M}_{1,\,{\rm wind}}, \\
&\dot{M}_2 = \dot{M}_{2,\,{\rm wind,\,accretion}} + \dot{M}_{2,\,{\rm wind,\,loss}}, \\
&\dot{M}_{2,\,{\rm wind,\,accretion}} = \beta_{\rm WRLOF} \dot{M}_{1,\,{\rm wind}}.
\end{align*}

The second panel of Figure \ref{fig:Mdot_t_wRLOF} shows the mass of the donor (blue solid line) and the accretor (orange solid line) as a function of time (the same time range is applied to all four panels). The third panel displays the accumulated mass from wind accretion ($\Delta M_\mathrm{2,\,wind,\,accretion}$; orange dashed line), wind mass loss ($\Delta M_\mathrm{2,\,wind,\,loss}$; dotted green line), and the total mass change of the accretor ($\Delta M_\mathrm{2}$; solid blue line). The last panel shows the surface rotation rate of the accretor (orange solid line), the critical rotation rate of the accretor (green dotted line), and the ratio between them (blue solid line).

This $M_1 = 1.33\,M_\odot$ donor star enters the AGB phase at 4.112 Gyr and has a wind loss rate of $\sim 10^{-10} - 10^{-9} \, M_{\odot}/{\rm yr}$. During the early stage of WRLOF (from 4.112 to 4.119 Gyr, a total of 7 Myr), the MT efficiency maxes out at 50\%, mainly due to the prescription by \citet{Abate2013} (as may also be seen in Equation \ref{wrlof}). This efficiency maintains at 50\% throughout the WRLOF phase. Compared to the late phase of the AGB evolution with a strong wind, the donor star's wind in the early AGB phase is relatively weak. Although the WRLOF efficiency is high, due to the low mass loss rate of the donor in this early phase, there is no significant change in the mass of both stars or in the surface rotation speed of the accretor (see the second and last panels of Figure \ref{fig:Mdot_t_wRLOF}). In addition, near 4.1183 Gyr, the donor star reaches the thermal pulse AGB phase, so the model switches to a stronger wind scheme (see Section \ref{sec:grid setting} for detailed information). The transition exhibits discontinuity rather than a smooth ramp. The duration of the dip is only 0.1 Myr. It is noteworthy that all other essential physical parameters continue to exhibit smooth behavior (e.g., mass, temperature, radius, luminosity). Therefore, this discontinuous change in the wind rate does not significantly impact the overall results. For smaller initial $q$ configurations with a more massive donor star (e.g. $M_1=3.3\,M_{\odot}$, same $M_2$), at the onset of the WRLOF, the WRLOF efficiency is $\sim 15\%$. We refer the reader to Section 2.2 and 3.3.2 of \citet{Abate2013} to check this efficiency for different parameter spaces.

From 4.119 Gyr to 4.1196 Gyr, the donor's wind becomes stronger and reaches about $10^{-6} \, M_{\odot}/{\rm yr}$ before the thermal pulse phase. During this 0.6 Myr period, the mass of the donor/accretor decreases/increases from 1.28/1.10 $M_{\odot}$ to 1.22/1.13 $M_{\odot}$, respectively (second panel of Figure \ref{fig:Mdot_t_wRLOF}). The accretor gains 0.03 $M_{\odot}$ in mass during this period (third panel). The accretor's surface rotation rate increases due to angular momentum transfer through the wind. Additionally, as $\omega_\mathrm{crit}\propto R^{-3/2}$, the accretor's radius expands from 1.3 $R_{\odot}$ to 1.7 $R_{\odot}$, resulting in a decrease in the critical rotation speed. The combination of the spin-up of the accretor and the drop in critical rotation speed causes $\omega_2/\omega_{2,{\rm crit}}$ to reach 1 at 4.1196 Gyr (last panel). At this moment, the boosted wind from the accretor turns on. In the discussion section \ref{subsec:Effects of Angular Momentum Transfer}, we compare a model with and without angular momentum transfer during the MT phase.

From 4.1196 Gyr until the end of the donor's AGB phase at 4.120 Gyr, the donor star undergoes several thermal pulses, causing its mass loss rate to vary between $10^{-6} \, M_{\odot}/{\rm yr}$ and $10^{-4} \, M_{\odot}/{\rm yr}$. The accretor maintains a critical rotation rate due to the WRLOF. The boosted wind is still activated to ensure that the accretor always rotates slightly below its critical rate (top panel, red dotted line of Figure \ref{fig:Mdot_t_wRLOF}). During this period, the donor star loses a significant amount of mass, decreasing from 1.22 to 0.55 $M_{\odot}$, while the accretor's mass only increases by 0.005 $M_{\odot}$. The third panel of Figure \ref{fig:Mdot_t_wRLOF} shows that the accretor gains 0.19 $M_{\odot}$ through WRLOF. However, the accreted material from the wind is nearly balanced out by the enhanced stellar wind loss from the accretor. 

In summary, our simulation confirms the significance of WRLOF during the early AGB phase when the wind speed is relatively slow. Once the accretor reaches the critical rotation rate and the boosted wind of the accretor removes the accreted material, the final mass of the accretor remains relatively unchanged. Overall, the accretor mass only increases by 3\%.

\begin{figure}[tp]
\centering
\includegraphics[scale=0.5,angle=0]{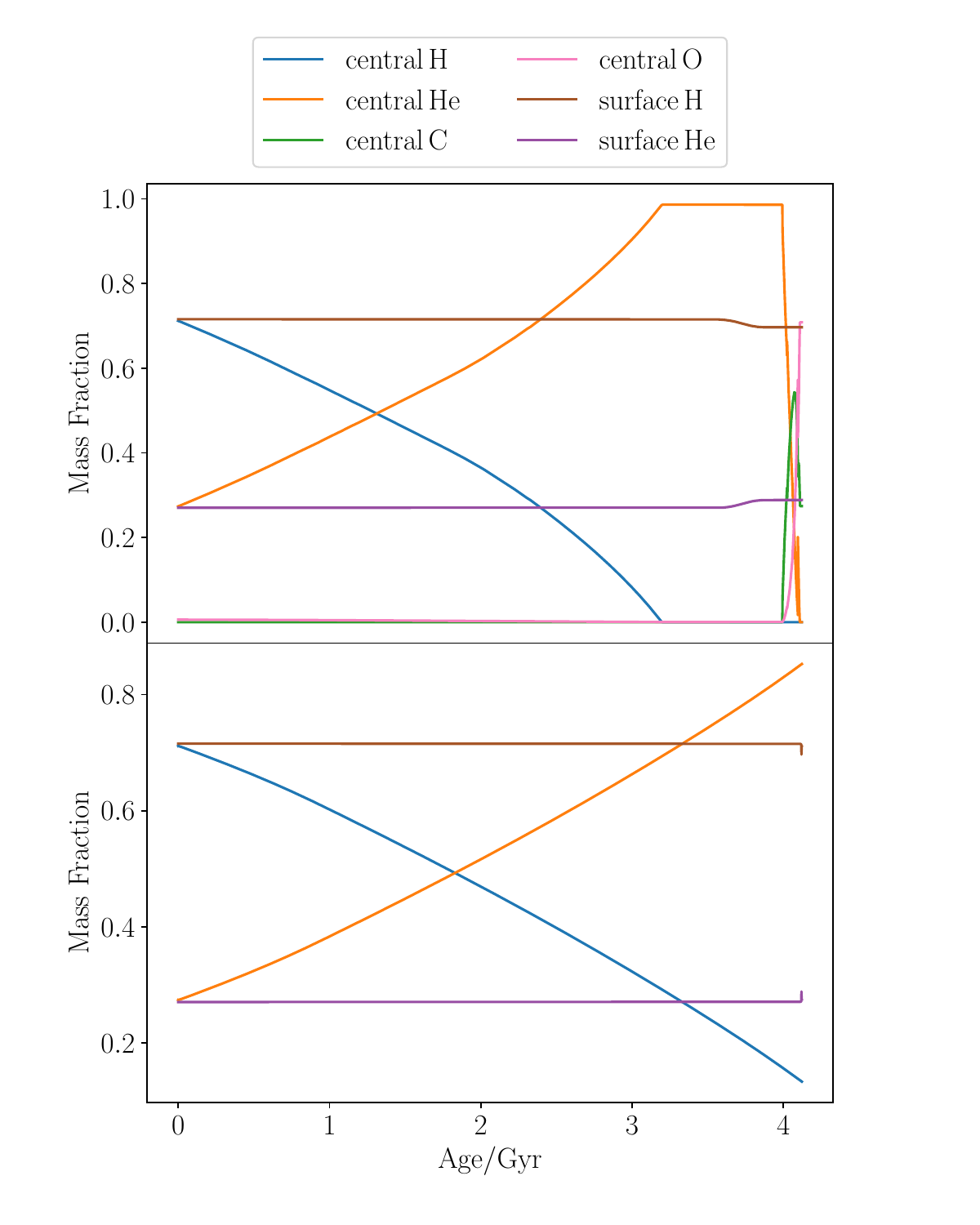} 
\caption{The chemical abundances of the donor (top panel) and the accretor (bottom panel), as a function of system age (in Gyr).}
\label{fig:all_abund_pf}
\end{figure}

In Figure \ref{fig:all_abund_pf} we show the chemical abundances as a function of time for both the donor (top panel) and the accretor (bottom panel). We display four abundances at the core: hydrogen (blue), helium (orange), carbon (green), and oxygen (pink). The surface hydrogen/helium is displayed in purple/brown, respectively. The donor star exhausts its central hydrogen at 3.2 Gyr, marking the end of the MS stage and the transition to its subgiant and RGB phase. The core helium is ignited at 4.0 Gyr, indicating the beginning of the horizontal branch, during which both the core carbon and oxygen abundances increase. The surface abundance of the donor star does not change significantly; the hydrogen mass fraction drops from 0.715 to 0.7, and the helium mass fraction increases from 0.27 to 0.29. During the WRLOF phase, all the elements in the donor star maintain constant mass fractions.

The accretor is still in its MS phase, during which it burns hydrogen into helium in the core. The surface hydrogen/helium mass fraction decreases/increases from 0.715/0.27 to 0.7/0.29, then recovers back to 0.71/0.275, respectively, due to the accretion of more helium-rich material, followed by thermohaline mixing. Thermohaline mixing is a process where differences in temperature and chemical composition lead to the mixing of material. In this context, as the star accretes helium-rich material, it undergoes thermohaline mixing, which homogenizes the composition at the stellar surface.

\section{Discussion}
\label{sec:discussion}

\subsection{Effects of Angular Momentum Transfer}
\label{subsec:Effects of Angular Momentum Transfer}

When MT occurs, material can be accreted onto the secondary star in two ways: the formation of an accretion disk (if the impact parameter of the stream is greater than the radius of the accretor) or the direct impact of the material onto its surface. The specific angular momentum gained by the accretor is different in each case and is discussed in detail in \citet{Lubow1975} and \citet{deMink2013}.

The treatment described in \citet{deMink2013} is included in our fiducial grids to model accretion onto the secondary (\texttt{do\_j\_accretion =.true.}). Additionally, we recognized that the angular momentum transfer outlined in \citet{Lubow1975} could serve as an upper limit for the case under investigation. For example, during MT onto the accretor, a portion of the angular momentum carried by the stream is transferred to the accretor, while another portion may be dissipated in the disk as a result of gas dynamics processes.

In this subsection, we discuss the effects of angular momentum transfer in binaries by running experimental grids where we ignore the angular momentum transfer, which allows more material from the donor star to be accreted onto the accretor in general. Most of the time, the accretor is far below its $\omega_\mathrm{2,\,crit}$.


\subsubsection{WRLOF without Angular Momentum Transfer and Stellar Rejuvenation}
\label{sec:WRLOF without AM Transfer}

Another grid of models was performed with the WRLOF prescription, using the same parameter coverage as described in Figure \ref{fig:m2}. The location of the different classes of MT islands is the same. The only difference between this model and the fiducial grid is that the accretor gains up to $\sim 0.4\,M_\odot$ in the ``WRLOF island". On average, the accretor gains over five times ($0.22\,M_\odot$) as much mass as in the models shown in Figure \ref{fig:m2}. The mass gained by the accretor is similar to that in a previous study \citep{Sun2023}, where, without considering angular momentum transfer and stellar rotation, the BSS as an accretor gains 0.19 $M_{\odot}$ from the donor's AGB wind.

The amount of mass that can be accreted onto the accretor in the fiducial grid is limited by whether the accretor can be spun up to a critical rotation rate, which triggers mass-loss via an enhanced wind from the accretor. This wind reduces the accreted material, resulting in only a small amount of mass being retained by the accretor. However, if we turn off the angular momentum transfer, the accretor's spin will not be accelerated at the onset of WRLOF, which allows more mass to be accreted.

\begin{figure}[tp]
\centering
\includegraphics[scale=0.47,angle=0]{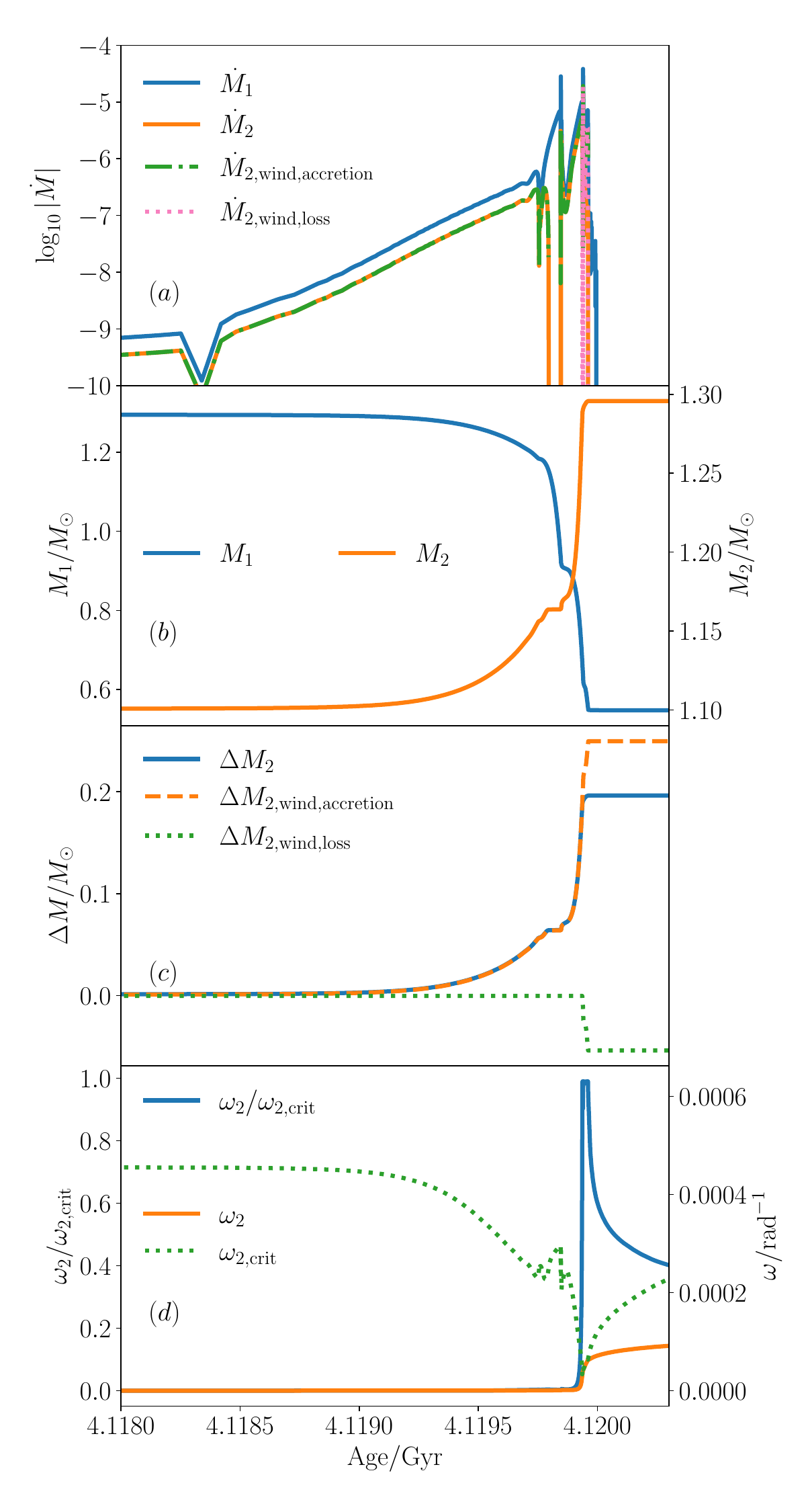} 
\caption{The detailed stellar and binary parameters are shown as a function of age in the angular momentum transfer turn-off model. The figure follows the description of Figure \ref{fig:Mdot_t_wRLOF}.}
\label{fig:nojdot_accr_Mdot_t}
\end{figure}

Similar to Figure \ref{fig:Mdot_t_wRLOF}, Figure \ref{fig:nojdot_accr_Mdot_t} shows the important binary evolution parameters as a function of time in the donor's AGB phase. This specific model has the same initial conditions as described in Figure \ref{fig:Mdot_t_wRLOF} to Figure \ref{fig:all_abund_pf}, where $M_1 = 1.33\,M_{\odot}$, $M_2 = 1.1\,M_{\odot}$ and $P_{\rm orb} = 3661$ days.

Compared to Figure \ref{fig:Mdot_t_wRLOF}, WRLOF lasts longer because the accretor does not spin up during MT. From the beginning of the AGB phase (4.11 Gyr) to 4.12 Gyr, a total of 10 Myr, $\dot{M}_2$ overlaps with $\dot{M}_\mathrm{2,\,wind,\,accretion}$ (top panel), during which the accretor gains mass from wind accretion. WRLOF efficiently increases the mass of the accretor from 1.1 $M_{\odot}$ to 1.28 $M_{\odot}$ in the second panel, with a mass gain of 0.18 $M_{\odot}$ from WRLOF accretion shown in the third panel. The WRLOF efficiency remains around 50\% for most of the time. As the radius of the accretor increases during the accretion, $\omega_\mathrm{2,\,crit}$ decreases until it reaches $\omega_2$ at 4.12 Gyr in the last panel, at which point the boosted wind from the accretor is developed, as shown by the red line in the top panel of Figure \ref{fig:nojdot_accr_Mdot_t}.

After the wind of the accretor is enhanced, the accretor increases in mass from 1.28 $M_{\odot}$ to 1.30 $M_{\odot}$, a change of 0.02 $M_{\odot}$, as shown in the second and third panels. As shown in the last panel, the accretor rotates close to its critical rotation rate at this stage. After the end of the donor's AGB phase, where wind accretion is no longer present, the accretor shrinks in radius from 6.5 $R_{\odot}$ to 1.5 $R_{\odot}$, resulting in an increase in both $\omega_2$ and $\omega_\mathrm{2,\,crit}$.

In summary, since no angular momentum is transferred to the accretor during the accretion process, the critical rotation rate of the accretor can only decrease through the expansion of the accretor at a later time. This allows WRLOF to efficiently transfer mass onto the accretor, resulting in a total mass gain that is five times greater than if angular momentum transfer had been considered in this low-mass binary example case.

\begin{figure}[tp]
\centering
\includegraphics[scale=0.5,angle=0]{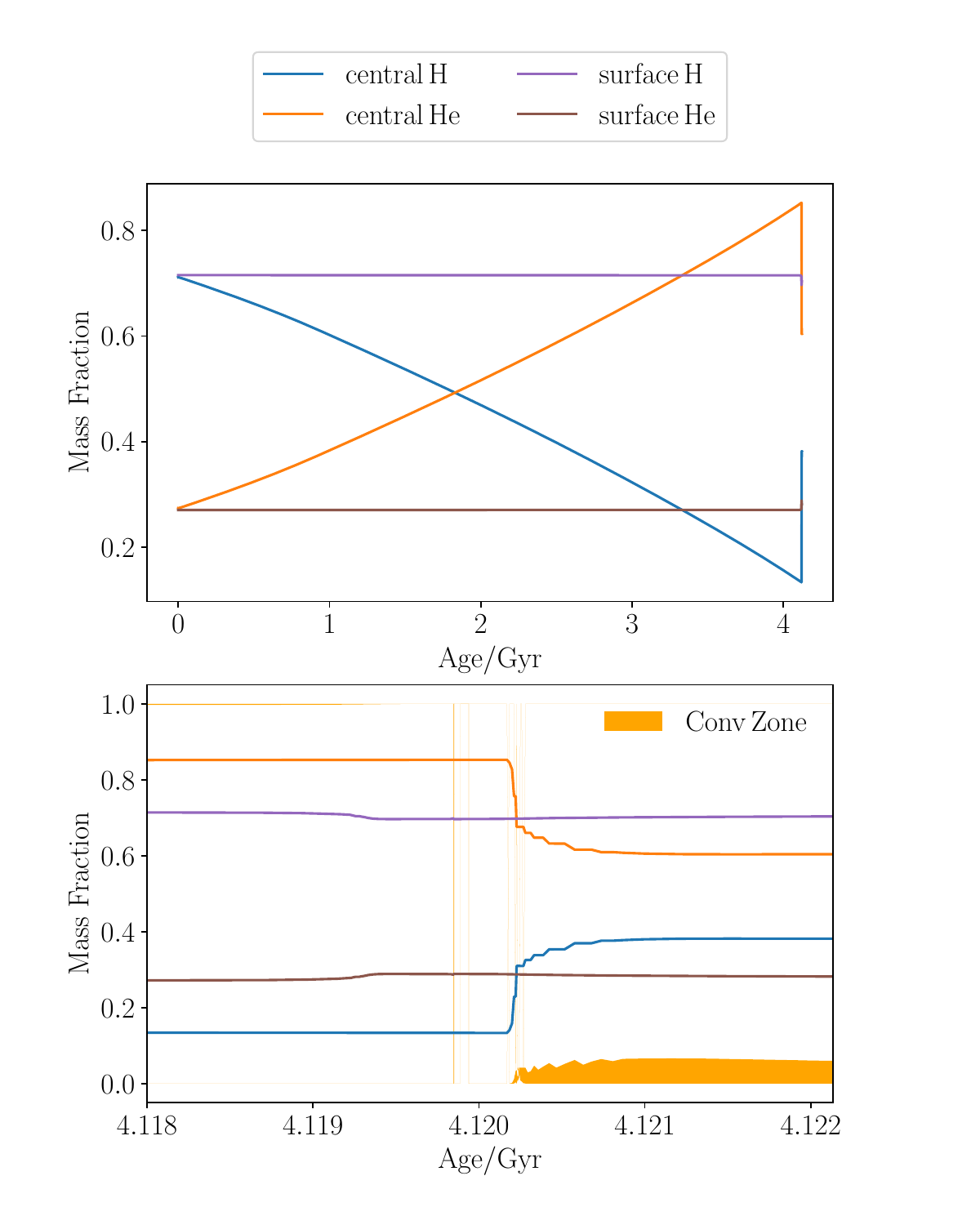} 
\caption{Top panel: the evolution of the central and surface hydrogen and helium abundances for the accretor over time (in Gyr). Bottom panel: the abundances and location of the largest convection zone during and after the WRLOF phase, over time. }
\label{fig:nojdot_accr_all_abund_pf}
\end{figure}

As the accretor gains mass from 1.1 $M_{\odot}$ to 1.3 $M_{\odot}$ through WRLOF, it undergoes rejuvenation. The top subplot of Figure \ref{fig:nojdot_accr_all_abund_pf} shows the evolution of central and surface hydrogen and helium abundances over time. The accretor burns central hydrogen into helium since it is in the MS phase. Due to WRLOF occurring at 4.11 Gyr, the central hydrogen abundance, shown by the blue solid line, increases, leading to a drop in central helium abundance. The bottom plot shows the abundance and location of the largest convection zone during and after WRLOF phase. The surface convection zone disappears when the accretor's mass reaches 1.3 $M_{\odot}$ at 4.12 Gyr. After that, the accretor develops a central convection zone, which brings more hydrogen fuel into the center. The central convection zone of the accretor is 0.08 $M_{\odot}$ after the WRLOF.

\subsubsection{BHL accretion without Angular Momentum Transfer}
\label{sec:discussion:bhl}

\begin{figure}[tp]
\centering
\includegraphics[scale=0.8,angle=0]{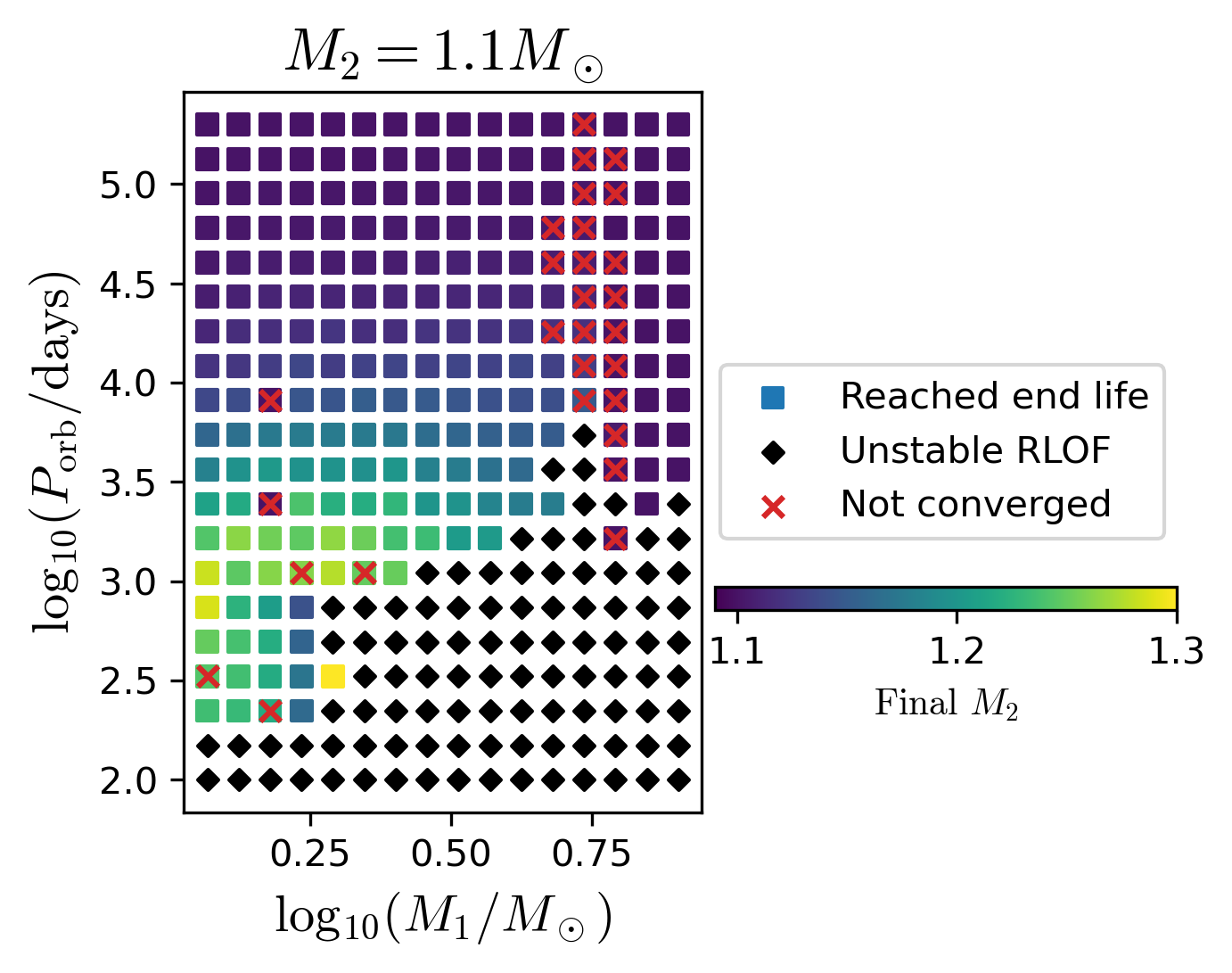} 
\caption{A slice of the grid with angular momentum accretion turned off and BHL accretion applied, with initial $M_2 = 1.1 M_\odot$. The marker shape shows the termination condition for each model and the color indicates the final accretor mass, $M_{\rm 2,f}$. The axes are the initial parameters for $M_1$ and $P_{\text{orb}}$ on a logarithmic scale. Analogous to Figure \ref{fig:m2}.}
\label{fig:nojdotm_bhl}
\end{figure}

To compare the results obtained with WRLOF and traditional BHL accretion, focusing on the same parameter coverage, the resulting grid with only BHL accretion (regardless of the wind speed compared with orbital speed) is shown in Figure \ref{fig:nojdotm_bhl}. For the BHL accretion efficiency, we used \texttt{MESA}'s default definition from \citet{Hurley2002}:
\begin{equation}
    \beta_{\text{BHL}} = \frac{\alpha}{2\sqrt{1-e^2}}\left(\frac{GM_2}{av_w^2}\right)^2 \left[1+\left(\frac{v_\text{orb}}{v_w}\right)^2\right]^{-\frac{3}{2}},
\label{bhl}
\end{equation}
where $\alpha = 1.5$ is a constant (the same as used in \citet{Abate2013}), $e$ is the eccentricity of the binary system, $G$ is the gravitational constant, and $a$ is the system separation.

The parameter range affected by BHL accretion is relatively small compared to the WRLOF island shown in Figure \ref{fig:end}. For lower mass systems with $M_1$ ranging from 1.17 to 2.3 $M_{\odot}$ and an initial $P_\mathrm{orb}$ of 1000 to 5500 days, BHL accretion can occur. For systems with an initial $M_1 > 2.3$ $M_{\odot}$, located in the region above the unstable MT island, BHL accretion can occur for an initial $P_\mathrm{orb}$ of 3000 to 10000 days. In contrast, the WRLOF island extends to initial $P_\mathrm{orb}$ of $10^5$ days. On average, systems undergoing BHL accretion gain mass of $\sim 0.1$ $M_{\odot}$, which is greater than the fiducial grid considering angular momentum transfer and the prescription for the boosted wind from the accretor. However, the average amount of material transferred from the BHL channel is less than the WRLOF ($\sim 0.2\,M_{\odot}$) channel without angular momentum transfer. In the BHL channel, relatively more material is accreted when the initial $P_\mathrm{orb}$ is around 1000 days, while wider initial $P_\mathrm{orb}$ results in a smaller amount of mass being accreted.

These results are based on models that do not consider angular momentum transfer during the MT and the prescription for the boosted wind from the accretor. When these two pieces of physics are combined, all types of MT prescriptions become less efficient, resulting in only a small amount of material being gained by the accretor.

\subsection{Fiducial model applied to Blue Lurker Observations}

Upon application of WRLOF, all fiducial models gaining $\gtrsim 0.01\,M_\odot$ are subject to critical rotation. Therefore, we seek to explore the implications of our findings on the unusually high rotation rates of BLs. Specifically, we examine how the fiducial models can explain (a) the location of BLs on the HR diagram as MS stars, (b) the origin of their rapid rotation, and (c) their present $P_\mathrm{orb}$.

We compare our results with the observational data of ten BLs with a possible late-phase companion in M67. Among these BLs, as shown in Table \ref{tab:blue lurkers}, five have $P_\mathrm{orb}<3000$ days, and the remaining five have $P_\mathrm{orb}>3000$ days. Table \ref{tab:blue lurkers} does not include WOCS 2068 in \citet{Leiner2019,Nine2023}, as this system has two MS stars that may not form through MT.

In the following paragraphs, we discuss how the models from the fiducial grid including regular RLOF can plausibly account for the data of the four close binaries ($<3000$ days) containing a BL, whereas the models from the WRLOF island can better explain the observations of BLs in wide binaries ($> 3000$ days). We only compared the model against the spectral energy distribution (SED) $T_\mathrm{eff}$ and the $\log\,g$ estimated from the \texttt{isochrones} \citep{Morton2015}. Although there is a lack of $\log\,g$ data for WOCS 4001, 11006, 1020, 2001 and 7035, they are still MS stars, so we assumed $\log\,(g/{\rm cm,s^{-2}}) = 4.25\pm 0.25$ with an estimated error bar, which roughly corresponds to the surface gravity range of a low-mass MS star. The other $\log\,g$ values in Table \ref{tab:blue lurkers} have a comparable error.


\begin{table}[]
\caption{Blue Lurker Data from \citet{Leiner2019,Nine2023}}
\centering
\begin{center}
\begin{tabular}{c c c c}
\hline\hline
WOCS ID	& $T_\text{eff}, \text{SED}\, (\rm K) $	& $\log_{10}(g/{\rm cm\,s^{-2}})$  & $P_{\text{orb}}$ (day)\\
\hline
4001$^{a}$ & $5580^{+60}_{-130}$ & - & 139.77\\
14020$^{b}$ & $5990^{+60}_{-110}$ & 4.45 & 358.9\\
12020$^{a}$ & $6190^{+100}_{-140}$ & 4.44 & 762\\
3001$^{b}$ & $6690^{+80}_{-160}$ & 4.29 &  128.14\\
9005$^{a}$ & $6500^{+90}_{-110}$ & 3.98 & 2769\\
6025$^{a}$ & $6200^{+150}_{-230}$ & 4.23 & 6265\\
11006$^{a}$ & $6540^{+210}_{-200}$ & - & $>3500$\\
1020$^{a}$ & 6840 & - & $>10,000^{c}$ \\
2001$^{a}$ & 6040 & - & $>10,000^{c}$ \\
7035$^{a}$ & 6470 & - & $>10,000^{c}$ \\
\hline
\multicolumn{4}{l}{

\begin{minipage}{9 cm}

\small $^{a}$ Note: 
Data from \citet{Leiner2019}.

\small $^{b}$ Note: 
Data from \citet{Nine2023}.

\small $^{c}$ Note: 
BLs with no significant RV variations, either in a wide system where $P_{\rm orb}>10^4$ days (more likely) or in a system observed nearly face-on. However, it is essential to note that the completeness of binary orbital solutions also depends on the orbital period, mass ratio, and eccentricity. For instance, in Figure 1 from \citet{Geller2012}, it shows that the completeness of the orbital solution drops significantly for long orbital periods and high eccentricities.
\end{minipage}
}
\end{tabular}
\end{center}
\label{tab:blue lurkers}
\end{table}

\begin{figure}[tp]
\includegraphics[scale=0.75,angle=0]{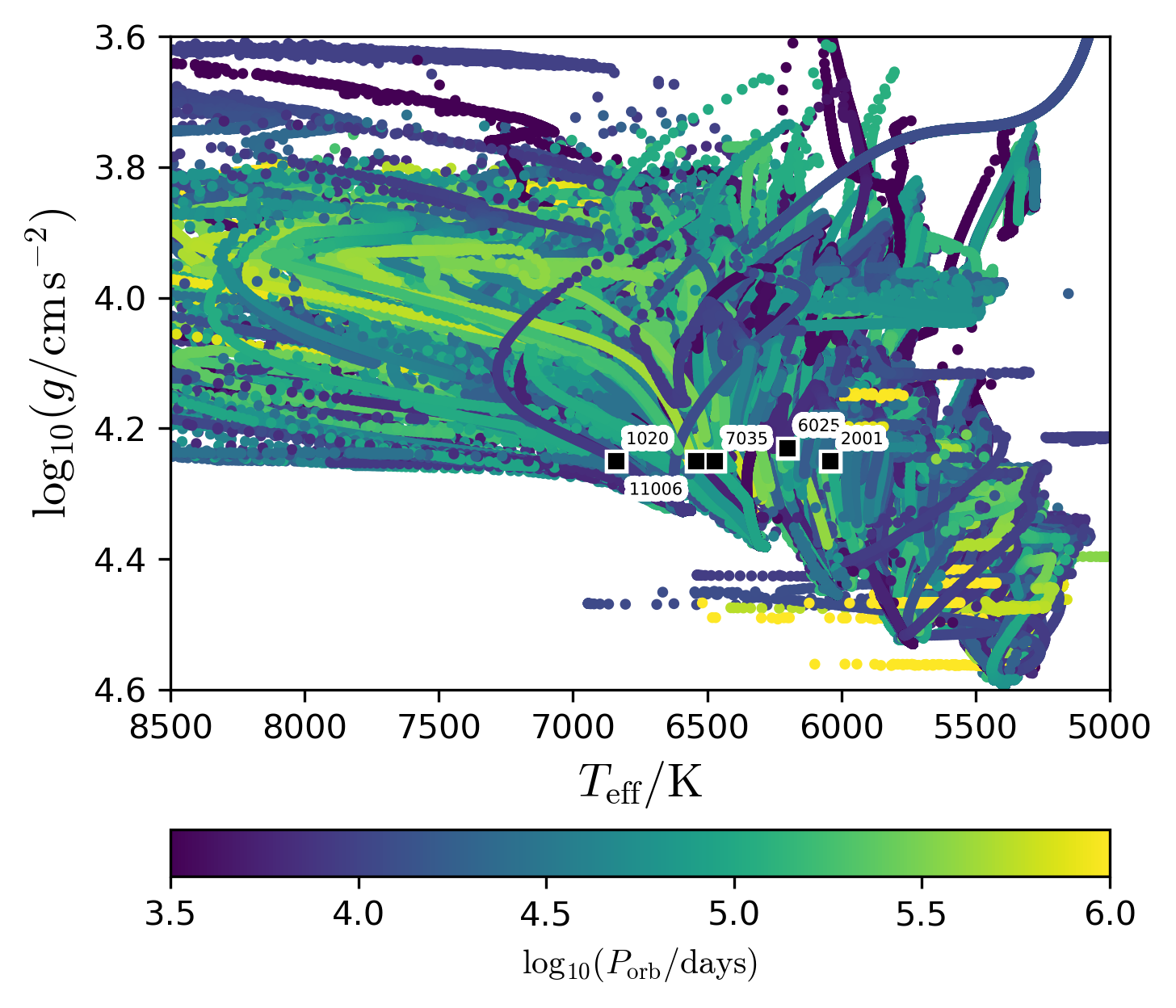} 
\caption{On the $\log\,g$ - $T_\mathrm{eff}$ plane, a plot showing the evolutionary tracks of the accretor for all models of the WRLOF channel from the fiducial grid described in Figure \ref{fig:end} to \ref{fig:rot}. BL data points with $P_{\text{orb}} > 3000$ days are shown in black, labeled with the WOCS ID.}
\label{fig:hrallwrlof}
\end{figure}

Figure \ref{fig:hrallwrlof} displays the colored evolutionary tracks of the accretor on the $\log\,g$ - $T_\mathrm{eff}$ plane, which are obtained from the converged models only in the ``WRLOF island" as well as a final $P_\mathrm{orb}> 3000$ days to explain the wide BL systems with $P_\mathrm{orb}> 3000$ days. The color of the tracks indicates the system's $P_\mathrm{orb}$ during the evolution. The tracks are selected with an accretor mass ranging from 0.9 to 1.3 $M_\odot$, but most of the models from the ``WRLOF island" have a final $P_\mathrm{orb}> 3000$ days. Due to the angular momentum transferred via MT, which tends to operate on a faster timescale than magnetic braking, which could otherwise counteract rapid rotation, the stars spin up to their critical rotation rate and develop enhanced winds, resulting in less efficient MT so that the stars can maintain their MS evolution. If the rotation of stars is not considered (see \ref{subsec:Effects of Angular Momentum Transfer} for more details) and material is allowed to keep accreting onto the accretor, WRLOF should be more efficient. The formation of wide orbit BLs is attributed to the combination of stars developing enhanced winds due to critical rotation treatment and WRLOF.


After a MT phase, magnetic braking could be important in spinning down the star. However, most of the BL data doesn't include the age since MT ceased, except for WOCS 3001 and 14020. Therefore, we did not attempt to match the model-predicted rotation period to the observed rotation period in this project. In this subsequent study \citep{2024arXiv240317279S}, we investigate how magnetic braking could account for the rotation periods of the post-MT accretors with ages measured from the end of MT. For WOCS 3001 and 14020, their rotation periods are 2 and 4.4 days, respectively. Indicated by their white dwarf cooling time, the MT for these systems ceased 300 to 900 Myr ago. Although the BLs from WOCS 3001 and 14020 are currently below critical rotation, they are still rotating much more rapidly than their single-star counterparts. Single stars typically spin down from the zero-age main sequence phase, with initial rotation rates usually ranging from 0.02 to 0.15 of the critical rotation rate \citep{2023ApJ...950...27G}. The key distinction when observing stellar spin-down from single-star evolution versus post-MT accretors is that the latter scenario provides an opportunity to test magnetic braking or other spin-down theories from the star's critical rotation speed. This study provides initial insights into explaining the origin of the rapid-rotating BLs.

The $P_{\rm orb}$ after WRLOF are still wide enough to match the data because most of the systems with WRLOF have an initial $P_\mathrm{orb}$ greater than 3000 days, and the wind loss through the binary system reduces the specific angular momentum while keeping the system expanding.



\begin{figure}[tp]
\includegraphics[scale=0.75,angle=0]{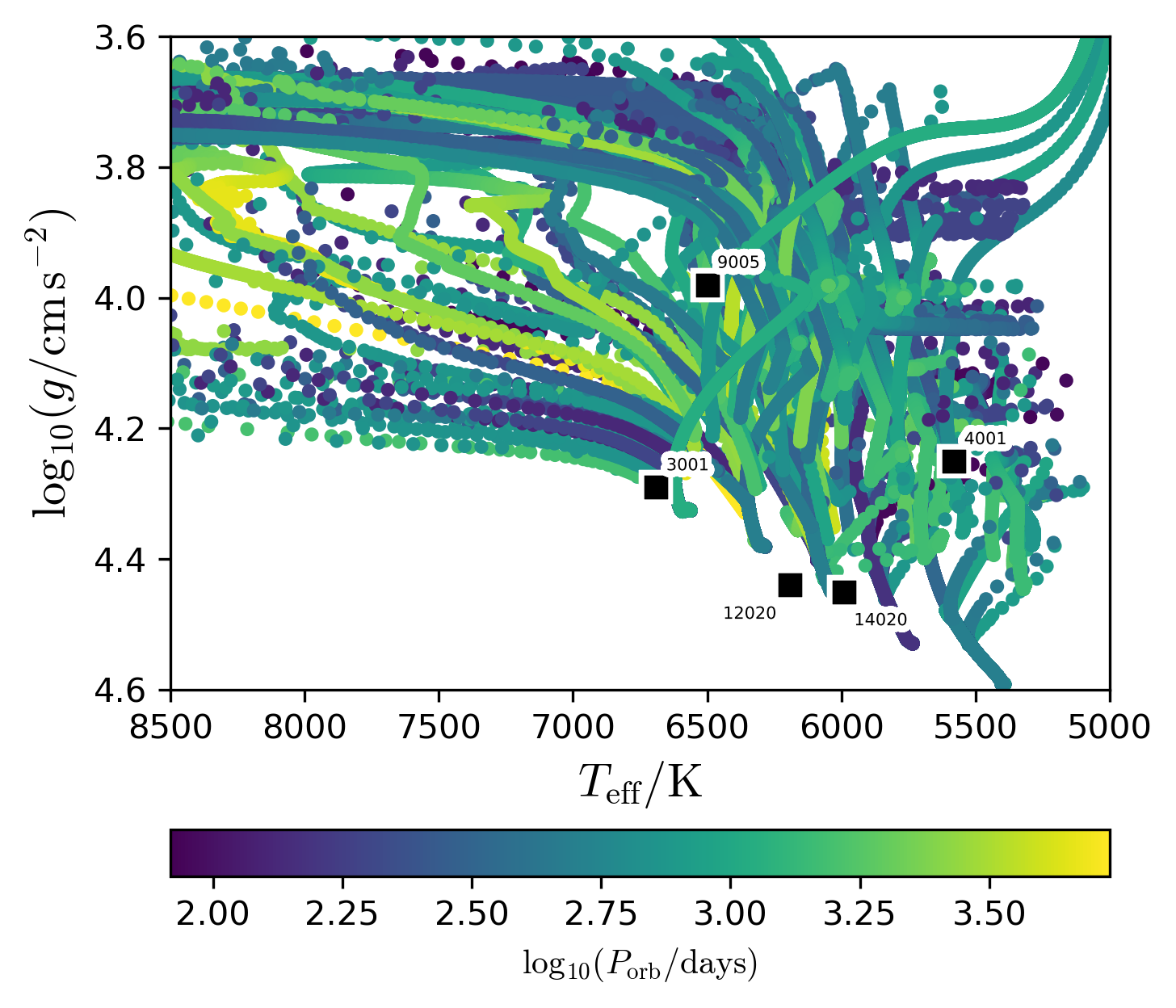} 
\caption{Analogous to Figure \ref{fig:hrallwrlof}, but the evolutionary tracks of the accretor for all models of the regular RLOF channel. BL data points with $P_{\text{orb}} < 3000$ days are shown in black and labeled with WOCS ID.}
\label{fig:hrlowprlof}
\end{figure}


Similar to Figure \ref{fig:hrallwrlof}, in Figure \ref{fig:hrlowprlof} we show the colored evolutionary tracks of the accretor from the regular RLOF channel, with a final $P_\mathrm{orb} < 3000$ days. The data for five BLs with $P_\mathrm{orb} < 3000$ days are shown as black squares. Due to the enhanced accretor wind, most of the accretors are still MS stars after regular RLOF, which qualitatively matches the data on the $\log\,g$ - $T_\mathrm{eff}$ plane. Accretors undergoing regular RLOF can also be spun up to near a critical rotation rate, and then the wind is boosted. After the case-B RLOF, most of the systems are still below $\sim 3000$ days, which qualitatively matches the five BLs with a relatively short orbital periods. The data (WOCS 3001, 12020 and 14020) in Figure \ref{fig:hrlowprlof} do not exactly overlap with the tracks. A set of fine-tuning models or models with an extended initial accretor mass (beyond our fiducial range of $0.9$ - $1.3$ $M_\odot$), initial metallicity, or mixing length factor range could better match the data.

\begin{figure}[tp]
\includegraphics[scale=0.48,angle=0]{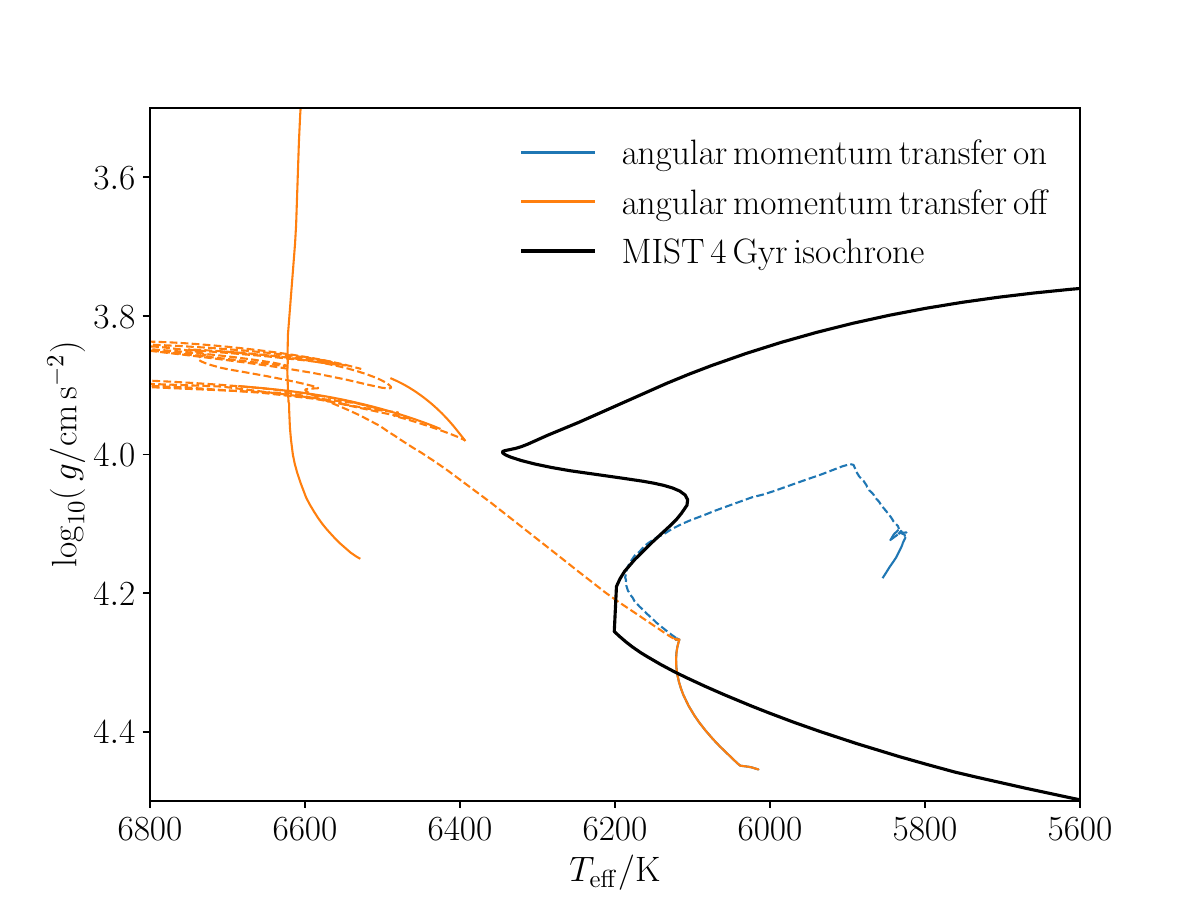} 
\caption{Two evolutionary tracks of the accretor are shown for the models described in Section \ref{sec:fiducial detailed model} (in blue) and Section \ref{sec:WRLOF without AM Transfer} (in orange) in the $\log\,g$ versus $T_{\rm eff}$ plane. These two models start with the same binary configuration: $M_1=1.33\,M_{\odot}$, $M_2=1.1\,M_{\odot}$, and $P_{\rm orb}=3661$ days. The model from the fiducial grid considers angular momentum transfer and the boosted wind mechanism, while the evolutionary tracks from Section \ref{sec:WRLOF without AM Transfer} do not take angular momentum transfer into account, resulting in the accretor gaining more mass from the donor. The line style is dashed when the MT rate exceeds $10^{-10}\,M_{\odot}/{\rm yr}$. The solid black line shows the 4 Gyr isochrone created by \texttt{MIST}, with M67 being around 4 Gyr.}
\label{fig:HR_accr_comb}
\end{figure}

One crucial factor in the formation of a BL versus a BSS is the accretion efficiency. Figure \ref{fig:HR_accr_comb} compares two accretor models: one from Section \ref{sec:fiducial detailed model} (represented by blue solid and dashed lines), and the other from Section \ref{sec:WRLOF without AM Transfer} (represented by orange solid and dashed lines), i.e. the RLOF+WRLOF grid with and without angular momentum transfer turned on during the MT, respectively.
Both models consider an example binary system with an initial accretor mass of 1.1 $M_{\odot}$. 
The evolution starts at ZAMS $(T_{\rm eff}/{\rm K},\, \log_{10}(g/{\rm cm\,s^{-2}}))=(6000, 4.45)$ and WRLOF occurs at $(T_{\rm eff}/{\rm K},\, \log_{10}(g/{\rm cm\,s^{-2}}))=(6100, 4.25)$. The dashed lines indicate a strong accretion rate, where $\dot{M}_a > 10^{-10}\,M_{\odot}/{\rm yr}$. For the fiducial model, $\dot{M}_a$ drops below $10^{-10}\,M_{\odot}/{\rm yr}$ at $(T_{\rm eff}/{\rm K},\, \log_{10}(g/{\rm cm,s^{-2}}))=(5825, 4.12)$, while for the model without angular momentum transfer, the strong accretion stops at $(T_{\rm eff}/{\rm K},\, \log_{10}(g/{\rm cm,s^{-2}}))=(6300, 3.15)$. The accretor running back and forth in the $\log\,g$ versus $T_{\rm eff}$ plane is due to AGB thermal pulses from the donor star causing variations in the MT rate.

When angular momentum transfer and boosted wind at $\omega/\omega_{\rm crit}\sim 1$ are considered, the fiducial model gains only 0.04 $M_{\odot}$, and its core region's structure and abundances remain unchanged. As a result, the star remains a MS star after WRLOF, located near its pre-WRLOF position on the $\log\,g$ versus $T_{\rm eff}$ plane. In contrast, if angular momentum were only partially transferred, or if an alternative efficient spin-down mechanism were present, more material could be transferred, leading to a rejuvenation of the accretor and its evolution into a BSS (moving beyond the MS turn-off, as the orange line shows). The old open cluster M67 has an age of $\sim 4$ Gyr \citep{Geller2021}. A 4 Gyr isochrone generated by the MESA Isochrones and Stellar Tracks code (\texttt{MIST}; \citealt{Dotter2016,Choi2016}) is shown in Figure \ref{fig:HR_accr_comb} as a reference point, using non-rotating and solar metallicity settings.


\subsection{Possible spin-down regime during the accretion}
\label{subsec:possible spin-down}
 
There are several BSSs confirmed to have long orbital periods of $\gtrsim 3000$ days. For instance, in the well-studied open clusters NGC 188 and NGC 6819, there are several BSSs with confirmed orbital period solutions or are confirmed as radial velocity (RV) variables (i.e. could have long orbital periods with further observations) \citep{Mathieu2009, Geller2012, Milliman2014}. Moreover, there are more than ten BSS systems that have not shown RV variability, and it is possible that some of them have extremely wide orbits where $P_\mathrm{orb} > 10^4$ days. After undergoing MT in binaries, BSSs evolve beyond the MS turnoff more massive than they were originally. Successfully acquiring this extra mass relies on  a more efficient MT process (i.e. an efficient spin down mechanism).

Some interaction between the incoming winds and the low-mass accretor's magnetosphere could regulate angular momentum transfer during MT, possibly allowing for more efficient MT. This interaction between an accreting, magnetically active star and its accretion flow has been studied in several scenarios. For example, the issue of how pre-main sequence (pMS) stars manage a roughly constant rotation rate during gravitational contraction and accretion requires an efficient angular momentum loss mechanism which may involve stellar winds \citep{matt.pudritz:2005,matt.pudritz:2008}, or magnetospheric ejections that may remove angular momentum in the presence of differential rotation between a star and its accretion disk \citep{romanova.etal:2009, zanni.ferrerira:2013}; e.g., \cite{pantolmos.etal:2020, ireland:2022} have performed some recent studies into the extent that these processes could play in pMS stars achieving spin equilibrium.

Whether similar processes may be at play in the case of a main sequence star accreting material from an AGB star (as in the present work) has been speculated on by \cite{shiber.etal:2016}, who suggest that main sequence stars could interact with their accretion disks to launch material, regulating angular momentum transfer. These processes have been explored further, especially in the context of a donor star embedded in a CE, and whether they may lead to CE ejection (see a recent review by \citealt{ropke.etal:2023}). Interactions between an accretion flow and the magnetosphere of low-mass main sequence stars could potentially impact the angular momentum transfer of these systems, and should be investigated further.

\section{Conclusions}
\label{sec:conclusions}

We incorporated the WRLOF prescription into the \texttt{MESA binary} module and explored the parameter space for low- and intermediate-mass stars in binaries. Both stars are evolved simultaneously. The donor's mass range was chosen to be from 0.9 to 8 $M_{\odot}$ to ensure that there is an AGB phase during the evolution, where the slow and dusty wind could potentially be transferred to the accretor. The initial $P_\mathrm{orb}$ ranged from 100 to $2 \times 10^5$ days, wide enough to allow for wind MT. The accretor's mass range was set from 0.9 to 1.3 $M_{\odot}$, with a focus on low-mass stars as possible progenitors of BLs. BLs are a newly discovered class of stars in the open cluster M67, which exhibit anomalous rapid rotation for their age during their MS evolutionary stage.

Our fiducial models with WRLOF and regular RLOF applied were able to provide a qualitative explanation for data of BLs. The inefficiency of MT and the effectiveness of angular momentum transfer in our model could account for the observed fast rotation of MS stars and their orbital periods.

In addition to the modeling, we summarize the astrophysical findings as follows:

\bigskip

1) In the fiducial grid of models, during MT, angular momentum is effectively transferred to the accretor, resulting in a spin-up of the accretor. At the same time, the accretor expands in radius, which lowers its critical rotation rate. The combination of increasing rotation speed and decreasing critical rotation rate causes $\omega/\omega_\mathrm{crit}$ to rapidly reach $\simeq 1$ after MT occurs. A boosted stellar wind is used to ensure that the accretor rotates slightly below the critical rotation rate, which reduces the incoming accretion material. In the fiducial model, the accretor gains $\simeq 1-5\%$ of its total mass after MT. This mass gain value is independent of the MT type (e.g., regular RLOF and WRLOF) and is less affected by the initial parameter space. The timescale of the WRLOF phase is equivalent to the donor's AGB phase.

Under similar physical assumptions, \citet{Schuemass2022} found a MT efficiency of less than 5\% in their study of binaries containing a B-type star. Similarly, in the \texttt{POSYDON} version one massive binary grids that focused on neutron stars and black hole progenitors, the accretors gained less than 5\% mass on average \citep{2024arXiv240307172A}.

This inefficient MT is crucial for explaining the BL data. If the accretors in the fiducial grid accrete too much material, they will evolve off the MS turn-off and become a BSS.


\bigskip

2) Stable regular RLOF happens when the system's initial $P_\mathrm{orb}$ is less than 1000 days and the mass ratio is close to 1. Unstable MT occurs when the initial $P_\mathrm{orb}< 1500$ days, resulting in a CE phase. WRLOF occurs at an initial $P_\mathrm{orb}$ ranging from $\sim 1700$ to $10^5$ days. The resulting $P_\mathrm{orb}$ after WRLOF ranges from 1424 to $1.954 \times 10^6$ days. This WRLOF mechanism could be the reason for a fast rotator in a wide orbit, from a post-MT system. 


\bigskip
3) If there were a mechanism to prevent the star from spinning up to near its critical rotation rate, such as halting the transfer of angular momentum during MT or ignoring the accretor's rotation rate, MT could be more effective than in our fiducial models. WRLOF is more efficient than traditional BHL accretion, especially in wide systems with an initial $P_\mathrm{orb}>10^4$ days. The accretor is minimally affected by BHL accretion (see Figure \ref{fig:nojdotm_bhl}), but could gain around $0.2\,M_{\odot}$ through WRLOF. Not all BL binaries contain information about the age since MT ceased (except for WOCS 3001 and 14020, see \citet{Nine2023}); hence, in this work we do not perform a systematic gyrochronology analysis to explain their current rotation rate and accurately assess their post-MT evolution. However, their current rapid rotation could originate from both WRLOF and regular RLOF.

Moreover, for a 1.1 $M_{\odot}$ accretor that has a radiative core, as it gains $\sim 0.2\,M_{\odot}$ in mass, the star could be rejuvenated, as the center of the 1.3 $M_{\odot}$ star is dominated by convective mixing, which brings hydrogen fuel from the shell to the core burning region. The star could become more luminous and hotter than the MS turn-off, as a BSS.

\bigskip

4) BLs with a $P_\mathrm{orb} < 3000$ days may result from regular RLOF. Regular RLOF in the fiducial grid has a $P_\mathrm{orb}$ from 377 to 5400 days after MT. 
On the other hand, BLs in wide binaries or those without significant RV variations may have evolved through the WRLOF.

\bigskip

In the grid with the angular momentum transfer turned off, a significant number of models experienced numerical convergence issues during the donor's AGB phase. At the tip of the thermal pulses, the AGB star loses mass through winds at a rate exceeding $10^{-3}\,M_{\odot}/\mathrm{yr}$. The behavior of the system is similar to that of a dynamically unstable MT associated with regular RLOF. One of the future aspects of this study is to perform higher-dimensional simulations and then discuss this ``wind CE" phase. \citet{Leiner2021} suggested that a more careful treatment of the wind MT in binary modeling needs to be considered to explain the population and distribution of the BSSs in old open clusters. Furthermore, for a given stellar cluster, the relative population between BSSs and BLs could help to constrain the low-mass binary modeling, as BSSs are more likely forming through more efficient MT while BLs are from systems with less efficient MT. Another future direction for this work is to conduct a population study with a more complete angular momentum evolution and MT efficiency prescription on those low-mass binaries.

\section*{acknowledgements}

This material is based upon work supported by the National Science Foundation (NSF) under Grant No. AST-2149425, a Research Experiences for Undergraduates (REU) grant awarded to CIERA at Northwestern University. Any opinions, findings, and conclusions or recommendations expressed in this material are those of the author(s) and do not necessarily reflect the views at the National Science Foundation. This research was supported in part through the computational resources and staff contributions provided for the Quest high performance computing facility at Northwestern University which is jointly supported by the Office of the Provost, the Office for Research, and Northwestern University Information Technology.

We thank the anonymous referee for providing numerous insightful comments that enhanced the clarity and depth of our research. M.S. and V.K. acknowledge the support from the GBMF8477 grant (PI Kalogera). M.S. thanks the entire \texttt{POSYDON} developer team for their invaluable technical support. M.S. also thanks Phil Arras, Xiaoshan Huang and Yoram Lithwick for insightful comments on angular momentum transfer in binaries. M.S. thanks Chenliang Huang for discussions from the inception of this idea to the final completion of this manuscript. S.G. acknowledge the funding support by a CIERA Postdoctoral Fellowship. A.M.G. acknowledges support from the NSF AAG grant No. AST-2107738. E.M.L. was supported in part by a CIERA Postdoctoral Fellowship.

\software{\texttt{numpy} \citep{2020NumPy-Array}, \texttt{scipy} \citep{2020SciPy-NMeth}, \texttt{pandas} \citep{mckinney2010data}, \texttt{Matplotlib} \citep{Hunter2007},\texttt{astropy} \citep{2013A&A...558A..33A,2018AJ....156..123A}, \texttt{MESA} 
\citep{Paxton2011,Paxton2013,Paxton2015,Paxton2018,Paxton2019,Jermyn2022}, \texttt{POSYDON} \citep{Fragos2022}.}

\bibliography{mypaper}
\bibliographystyle{aasjournal}

\end{CJK*}
\end{document}